\documentclass[12pt]{article}
\usepackage[latin1]{inputenc}

\usepackage{lipsum}
\usepackage{amsmath}
\usepackage{color}
\usepackage{amsfonts}
\usepackage[mathscr]{euscript}
\usepackage{graphicx}
\usepackage{geometry}
\usepackage{amssymb,epsfig}
\usepackage{subfigure}
\usepackage{comment}
\usepackage{tabularx}
\usepackage{bm}
\usepackage[dvipsnames]{xcolor}
\usepackage{exscale}
\usepackage{amsbsy}
\usepackage{textcomp}
\usepackage{hyperref}
\usepackage{slashed}
\usepackage{authblk}
\usepackage{tabularx}
\usepackage{color}
\usepackage{exscale}
\usepackage{doi}
\usepackage{tensor}
\usepackage{wrapfig}
\usepackage[font=footnotesize,labelfont=bf]{caption}
\usepackage{ulem} 
\usepackage{tikz-feynman}
\tikzfeynmanset{compat=1.1.0}
\usepackage{slashed}

\usepackage{makecell}

\def\ba#1\ea{\begin{align}#1\end{align}}
\def\bg#1\eg{\begin{gather}#1\end{gather}}
\def\bm#1\em{\begin{multline}#1\end{multline}}
\def\bmd#1\emd{\begin{multlined}#1\end{multlined}}

\newcommand{\be}{\begin{equation}}
	\newcommand{\ee}{\end{equation}}
\newcommand{\bea}{\begin{eqnarray}}
	\newcommand{\eea}{\end{eqnarray}}

\newcommand{\matleft}{\left(\begin{array}}
	\newcommand{\matright}{\end{array}\right)}
\newcommand{\Tr}{\operatorname{Tr}}

\usetikzlibrary{arrows.meta,bending}

\usepackage{calligra}
\DeclareMathAlphabet{\mathcalligra}{T1}{calligra}{m}{n}
\DeclareFontShape{T1}{calligra}{m}{n}{<->s*[2.2]callig15}{}
\newcommand{\scriptr}{\mathcalligra{r}\,}

\usepackage[numbers,sort&compress]{natbib}
\def\simge{
	\mathrel{\rlap{\raise 0.511ex 
			\hbox{$>$}}{\lower 0.511ex \hbox{$\sim$}}}}

\def\simle{
	\mathrel{\rlap{\raise 0.511ex 
			\hbox{$<$}}{\lower 0.511ex \hbox{$\sim$}}}}

\makeatletter
\renewcommand\section{\@startsection {section}{1}{\z@}%
	{-3.5ex \@plus -1ex \@minus -.2ex}
	{2.3ex \@plus.2ex}%
	{\normalfont\large\bfseries}}
\renewcommand\subsection{\@startsection{subsection}{2}{\z@}%
	{-3.25ex\@plus -1ex \@minus -.2ex}%
	{1.5ex \@plus .2ex}%
	{\normalfont\bfseries}}
\renewcommand\subsubsection{\@startsection{subsubsection}{3}{\z@}%
	{-3.25ex\@plus -1ex \@minus -.2ex}%
	{1.5ex \@plus .2ex}%
	{\normalfont\itshape}}
\makeatother

\def\pplogo{\vbox{\kern-\headheight\kern -29pt
		\halign{##&##\hfil\cr&{\ppnumber}\cr\rule{0pt}{2.5ex}&\ppdate\cr}}}
\makeatletter
\def\ps@firstpage{\ps@empty \def\@oddhead{\hss\pplogo}%
	\let\@evenhead\@oddhead 
}
\hypersetup{
	unicode=false,          
	pdftoolbar=true,        
	pdfmenubar=true,        
	pdffitwindow=false,     
	pdfstartview={FitH},    
	pdftitle={Boundary issues},    
	pdfauthor={},     
	pdfsubject={Subject},   
	pdfcreator={},   
	pdfproducer={}, 
	pdfkeywords={keyword1} {key2} {key3}, 
	pdfnewwindow=true,      
	colorlinks=true,       
	linkcolor=OliveGreen, 
	citecolor=NavyBlue,        
	filecolor=magneta,      
	urlcolor=cyan           
}

\numberwithin{equation}{section}

\newcommand*\samethanks[1][\value{footnote}]{\footnotemark}

\newcommand\beal{\begin{equation}\begin{aligned}}
		\newcommand\eeal{\end{aligned}\end{equation}}

\begin{document}

\normalem

\setcounter{page}0
\def\ppnumber{\vbox{\baselineskip14pt
}}

\def\ppdate{
} 
\date{}

\title{\LARGE\bf Extraordinary boundary correlations \\ at deconfined quantum critical points}
\author{Hao-Ran Cui and Hart Goldman}
\affil{\it\small School of Physics and Astronomy, University of Minnesota, Minneapolis, MN 55455, USA}
\maketitle\thispagestyle{firstpage}

\begin{abstract}

Recent years have seen a growing appreciation for the effects of quantum critical fluctuations on gapless boundary degrees of freedom. Here we consider the boundary dynamics of the non-compact $\mathbb{CP}^{N-1}$ (NCCP$^{N-1}$) model in two spatial dimensions, with $N$ complex boson species coupled to a fluctuating $\mathrm{U}(1)$ gauge field. These models describe quantum phase transitions beyond the Landau paradigm, such as the deconfined quantum critical point between superconducting (SC) and quantum spin Hall (QSH) phases. We show that, in a large-$N$ limit and with the bulk tuned to criticality, boundaries of the NCCP$^{N-1}$ model display logarithmically decaying, or ``extraordinary-log,'' correlations. In particular, when monopole operators exhibit quasi-long-ranged order at the boundary, we find  that the extraordinary-log exponent of the NCCP$^{N-1}$ model in the large-$N$ limit is $q=N/4$, signifying a new family of boundary universality classes parameterized by $N$. In the context of the QSH~--~SC transition, the quantum critical point inherits helical edge modes from the QSH phase, and this extraordinary-log behavior manifests in their Cooper pair correlations.

\end{abstract}

\pagebreak
\tableofcontents 
\pagebreak 

\section{Introduction}

Gapped topological phases of matter quintessentially exhibit protected boundary states embodying universal bulk responses. For example, fractional quantum Hall phases support chiral edge modes possessing fractional electromagnetic charge and exchange statistics. These properties are in turn imprinted on boundary observables. Prominent examples include universal scaling behavior~\cite{Chamon1995,chamon1996electronic,Sandler_1998,wen2004quantum} in shot noise~\cite{depicciotto1997,dolev2008, Hashisaka2015,Veillon:2024qkw} and  tunnelling~\cite{Chang1996,Cohen_2023,Manfra2025,Manfra2025a} at quantum point contacts. When topological phases are brought close to quantum critical points, however, a number of basic questions arise: To what extent do the edge modes persist at the quantum critical point? Do the bulk critical fluctuations alter boundary correlations in a universal, observable way? What kinds of boundary phenomena are accessible to gapless systems which would be prohibited for gapped topological phases? These questions have been partially addressed through the theory of gapless symmetry protected topological (SPT) phases~\cite{Keselman2015,jiang2017symmetryprotectedtopologicalluttinger,Scaffidi2017,Parker2018,Verresen2018,Verresen:2020qqf,Thorngren2021,Potter2023}, but developing a comprehensive understanding of boundary quantum critical phenomena remains an open challenge.  

Recently, in the context of the O$(n)$ model, it was discovered that degrees of freedom living on the 1$d$ boundaries of 2$d$ quantum critical systems can display correlations subverting traditional expectations for quasi-long-ranged order~\cite{Metlitski:2020cqy}. Dubbed ``extraordinary-log'' correlations, the boundary order parameter fluctuations decay as ${G(\rho)\sim (\log|\rho|)^{-q}}$, where $\rho$ is a boundary spacetime coordinate and $q$ is a new universal exponent. This behavior contrasts with the power law correlations associated with 1$d$ quantum (or 2$d$ classical) systems exhibiting quasi-long-ranged order. Although boundary critical phenomena in O$(n)$ models is an old and well studied subject~\cite{mills1971surface,binder1972phase,binder1974surface,Bray:1977tk,Bray:1977fvl,Ohno:1983lma,10.1143/PTP.72.736,McAvity:1995zd,Diehl:1996kd}, that such novel behavior could occur in 2$d$ went  unnoticed. Subsequent work has thus sought to verify extraordinary-log behavior numerically using Monte Carlo~\cite{Toldin:2020wbn,hu2021extraordinary,Toldin:2021kun,Sun:2023vwy,Toldin:2024pqi} and conformal bootstrap methods~\cite{Padayasi:2021sik}, as well as to extend the result to new contexts sharing the same O$(n)$ universality classes~\cite{sun2022quantum,Krishnan:2023cff,lee2023quantum,Cuomo:2023qvp,toldin2025extraordinary,sun2025boundary}.

In this work, we explore the possibility that extraordinary-log correlations can afflict topologically protected edge modes as the bulk system experiences a quantum phase transition. We focus on the deconfined quantum critical point (DQCP)~\cite{senthil2004deconfined,Senthil2004PRB,Senthil:2023vqd} between an $s$-wave superconductor (SC) and a quantum spin Hall (QSH) phase supporting helical edge modes~\cite{Grover2008,Liu:2018sww}. If continuous, such phase transitions are beyond the Landau paradigm, as SU$(2)$ spin rotation symmetry is spontaneously broken in the QSH phase, while U$(1)$ electromagnetic (EM) charge conservation is spontaneously broken in the SC. There is some phenomenological motivation for studying such transitions: Doping-tuned evolution between QSH and SC phases has been observed in WTe$_2$ monolayers~\cite{Song_2024,song2025unconventionalsuperconductingphasediagram}.

A low-energy effective field theory describing the bulk QSH -- SC transition is the non-compact $\mathbb{CP}^{1}$ (NCCP$^{1}$) model~\cite{Grover2008}, where two species of complex scalar boson strongly interact through an emergent U$(1)$ gauge field. The monopoles of this gauge field are conserved and interpreted as SU$(2)$ spin skyrmions carrying EM charge-2$e$. Although there are no gapless fermions in the bulk, the edge of the system hosts helical fermions, which are remnants of the QSH phase that couple to the critical bulk through gauge fluctuations~\cite{Thorngren2021,Ma:2021dfx,Potter2023}. Away from criticality, when the bulk scalar fields condense, the SU$(2)$ spin rotation symmetry is broken, the bulk becomes insulating, and the boundary fermions become the usual QSH edge modes. When the bulk scalars are gapped, strong gauge fluctuations bind the boundary fermions into Cooper pairs, and monopoles -- the charge-2$e$ skyrmions -- condense in the bulk, leading to superconductivity.

We consider a generalization of the bulk model to $N$ complex boson species, called the NCCP$^{N-1}$ model, and we retain the boundary fermions. This is equivalent to promoting the SU$(2)$ global spin rotation symmetry to SU$(N)$, and it  enables us to develop a controlled large-$N$ expansion for the model. The large-$N$ limit also heads off the possibility of a weakly first-order transition, as there is ample evidence that the NCCP$^1$ model is either weakly first order~\cite{Chen:2013ap,Nahum:2015jya,Nakayama:2016jhq,Poland:2018epd,Zhou2024} or multicritical~\cite{Takahashi:2024xxd}.

\begin{figure}[t]
    \centering
    \includegraphics[width=0.6\linewidth]{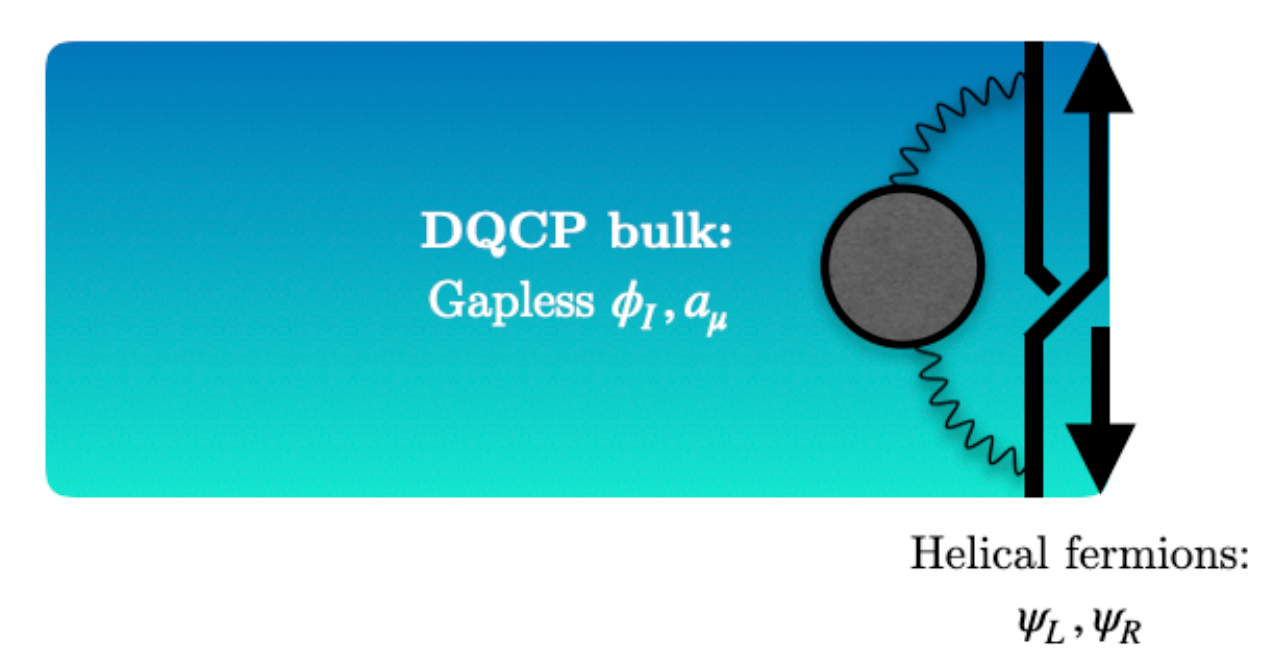}
    \caption{We consider a bulk system tuned to the QSH -- SC DQCP, which is described by a model of $N$ complex boson species, $\phi_I$, coupled to an emergent U$(1)$ gauge field, $a_\mu$. At the boundary, charged helical edge modes $\psi_L,\psi_R$ are inherited from the QSH phase. The bulk matter is gapped at the boundary, but gauge fluctuations  still couple to the boundary fermions. The critical bulk matter screens the emergent photon fluctuations (solid bubble), engendering extraordinary-log correlations of the boundary SC order parameter.}
    \label{fig: schematic}
\end{figure}

To obtain extraordinary-log correlations, we focus on the situation where the bulk is at criticality but the boundary is biased toward SC order.
This scenario corresponds to a particular choice of boundary conditions where the helical fermions remain coupled to the emergent bulk photon, while bulk matter fields become gapped at the boundary (see Fig.~\ref{fig: schematic}). The bulk critical matter screens the photon fluctuations, leading to extraordinary-log correlations among the boundary fields after integrating out the bulk degrees of freedom. 

In the large-$N$ limit, we compute the extraordinary-log exponent of the SC order parameter on the boundary, $\Delta_{\mathrm{SC}}$,
\begin{align}
\Big\langle \Delta_{\mathrm{SC}}(\rho)\,\Delta_{\mathrm{SC}}(0)\Big\rangle\sim\frac{1}{(\log\rho)^q}\,,\qquad q=N\left(\frac{1}{4}+\mathcal{O}(N^{-1})\right)\,.
\end{align}
Our result establishes a new family of extraordinary-log universality classes parameterized by $N$, which could be tested numerically. The large-$N$ approach we develop here can in principle be extended well beyond the NCCP$^{N-1}$ model to boundaries of gapless systems strongly coupled to gauge fields, including cases involving bulk Fermi surfaces. We hope our techniques may be leveraged in the future to make predictions relevant to experiments, for example for interfaces between quantum critical systems and superconductors.

We remark that the possibility of extraordinary-log behavior in an easy-plane version of the NCCP$^{N-1}$ model was anticipated in Ref.~\cite{myerson2024pristine}, although the exponent $q$ was not computed. Our work corroborates this proposal through explicit calculation, developing a large-$N$ framework for boundary criticality in gauge theories along the way. Additionally, Ref.~\cite{myerson2024pristine} argued that \emph{any} choice of boundary for the easy-plane model should host operators with extraordinary-log correlations, as a manifestation of the intertwinement of the orders on either side of the phase transition. We discuss how this prediction extends to the SU$(N)$-invariant version of the model. In particular, we propose that for $N$ larger than some critical value, the boundary we study corresponds to the \emph{only} stable, symmetry-preserving boundary condition, in contrast to the easy-plane case.

We proceed as follows. In Section~\ref{sec. boundary ward identity}, we present a high-level discussion of where extraordinary-log correlations come from, based on how bulk conservation laws are deformed in the presence of a boundary. In Section~\ref{sec: extra-ordinary NCCP$^{N-1}$ model}, we introduce the NCCP$^{N-1}$ model and define its extraordinary boundary conditions. In Section~\ref{sec: calculation}, we develop our large-$N$ expansion and present the calculation of the universal extraordinary-log exponent, $q$. We conclude in Section~\ref{sec: discussion}. A review of boundary criticality in the O$(n)$ model is presented in Appendix~\ref{app: review of o(n) boundary criticality}, and the details of our large-$N$ calculation, along with a complete solution to the large-$N$ saddle point equations for the gauge correlators, are given Appendix~\ref{app: extra-ordinary-log ah}. 

\section{Extraordinary-log correlations from global conservation laws}
\label{sec. boundary ward identity}
\subsection{Boundary Ward identities}

The possibility of fixed points with extraordinary-log correlations can be understood based on the deformation of conservation laws in the presence of a boundary, as established in Refs.~\cite{Padayasi:2021sik,Cuomo:2023qvp}. To set the stage for our analysis, we review these ideas in a manner that can easily be adapted to study boundaries in the NCCP$^{N-1}$ model.  

We start with a general model in $D=3$ Euclidean spacetime dimensions, which we take to be invariant under a U$(1)$ global symmetry (although the discussion here can be generalized to any continuous symmetry). Without a boundary, U$(1)$ invariance leads to the classical conservation law for its current, $J_\mu$, $\mu=\tau,x,y$, 
\begin{align}
    \partial^{\mu}J_{\mu}=0\,,
    \label{eq: current conservation on operator level}
\end{align}
which when treated as an operator equation gives rise to Ward identities.  

Introducing a boundary to the system alters bulk conservation laws like Eq.~\eqref{eq: current conservation on operator level} to account for the possibility of current fluctuations normal to the boundary,   
\begin{align}
\label{eq: boundary Ward general}
\partial_\mu J^\mu(r)=\delta(y)\,t(\rho)\,,
\end{align}
where we have placed the boundary at the $y=0$ plane and consider the bulk region to be $y>0$. We use $r=(\tau,x,y)$ as bulk coordinates, and $\rho=(\tau,x)$ denotes the coordinates in the plane of the boundary. The boundary operator, $t(\rho)$, is typically referred to as the ``tilt,'' and we will see below it has a special status in operator product expansions of bulk fields as the boundary is approached~\cite{Metlitski:2020cqy,Padayasi:2021sik,Cuomo:2023qvp}. Integrating both sides of Eq.~\eqref{eq: boundary Ward general} over a Gaussan pillbox surface enclosing the boundary implies the operator identity, 
\begin{align}
J^y(\rho,y\rightarrow0) = t(\rho)\,,
\end{align}
which tells us that the tilt depends on the boundary conditions placed on the normal component of the current, $J^y$.

Unsurprisingly, a generic current profile on the boundary explicitly breaks the global symmetry, as charge could be permitted to leave the system. To preserve charge conservation, 
it is sufficient to require that the tilt be a total derivative on the boundary~\cite{Thorngren2020boundary}, 
\begin{align}
\label{eq: tilt total derivative}
t(\rho)=\partial_m v^m(\rho)\,,
\end{align}
for some vector field, $v^m(r)$, $m=\tau,x$, which may involve degrees of freedom living exclusively on the boundary. In more modern language, this condition ensures the existence of a topological operator that is conserved across the whole system. Hence, the correct conserved charge typically has bulk and boundary contributions, ${Q=i\int dxdy[J^\tau+v^\tau\delta(y)]}$.

The operator equations \eqref{eq: boundary Ward general} -- \eqref{eq: tilt total derivative} are valid constraints for any symmetry-preserving boundary. If a disordered, gapped boundary is possible, then we may simply fix boundary conditions where the tilt vanishes, $J^y=t=0$. In other words, no charge is permitted to flow across the boundary. 

By contrast, ordered boundaries possess gapless ``spin wave'' degrees of freedom, causing $J^y$ to fluctuate on the boundary. In this case, the total current integrated across the boundary nonetheless vanishes by Eq.~\eqref{eq: tilt total derivative}, ensuring that although the current density may fluctuate, no net charge exits the system. For boundaries with spacetime dimension $D>2$, these boundaries spontaneously break\footnote{Throughout this work, we use the term \emph{order} broadly to also include cases of quasi-long-ranged and extraordinary-log order parameter correlations. We  reserve the phrase \emph{spontaneous symmetry breaking} for cases of long-ranged order.} the symmetry, and the boundary spin waves may be interpreted as genuine Goldstone modes localized to the boundary. For the cases of interest to us: two-dimensional boundaries of $D=3$ (spacetime) dimensional systems, long-ranged order is not expected, with the boundary order parameter fluctuations instead exhibiting either power law or extraordinary-log correlations. 

Gapless boundary degrees of freedom are also necessary if the system experiences an emergent 't~Hooft anomaly, which precludes the possibility of a trivially gapped, symmetry-preserving boundary~\cite{Thorngren2020boundary}. For example, in the QSH effect, helical boundary fermions are necessary to preserve charge and spin-$S_z$ conservation symmetries, which have a mixed $\mathrm{U}(1)_{\mathrm{EM}}\times\mathrm{U}(1)_{z}$ anomaly. As the bulk of such a system is tuned to a QSH -- SC DQCP with fully restored SU$(2)$ spin rotation symmetry, the boundary fermions remain to neutralize the DQCP's own emergent\footnote{Strictly speaking, the QSH -- SC transition is not anomalous, due to the fact that it possesses microscopic, gapped fermions in the bulk. Because these fermions are gapped and invisible to the bulk critical fluctuations, the anomaly is said to be emergent, in the sense that it still constrains the possible boundary terminations of the theory. In the context of the more commonly discussed N\'{e}el -- VBS DQCP in square lattice Heisenberg antiferromagnets, this anomaly is only ameliorated by the fact that the VBS symmetry is not on-site.} 't~Hooft anomaly~\cite{Thorngren2021,Potter2023}, and they couple to the bulk critical fluctuations~\cite{Ma:2021dfx,myerson2024pristine}. For the ordered boundary condition of the DQCP we will focus on, the aforementioned ``spin waves'' will simply be Cooper pair fluctuations of the boundary helical fermions. Consequently, we will find that not only do the helical fermions persist at the quantum critical point, but their correlations will exhibit the same extraordinary-log behavior discovered in Ref.~\cite{Metlitski:2020cqy} for the boundary spin waves in the O$(2)$ model.

\subsection{Effective action for extraordinary boundaries}

We are interested in quantum critical systems supporting gapless degrees of freedom confined to a boundary, where the combined bulk-boundary system preserves a global U$(1)$ symmetry. If the bulk and boundary remain coupled at low energies, these systems are said to be in an \emph{extraordinary} universality class. In the case of the O$(2)$ model, these boundary degrees of freedom result from a situation where boundary spins tend to remain ordered at the critical point, leading to boundary spin waves that can naturally be described in terms of a compact boson field on the boundary. On the other hand, we alluded above to situations like the QSH -- SC transition, where edge modes are inherited from the QSH phase. Here as well, bosonization enables a description of the edge modes in terms of a compact boson that can couple to bulk degrees of freedom in nearly the same way. We now turn to explain how the Ward identities developed above can motivate an effective theory germane to both contexts, which will clarify how extraordinary-log correlations arise. In Section~\ref{sec: extra-ordinary NCCP$^{N-1}$ model}, we will elaborate on the interpretation of this theory for the QSH -- SC transition. 

We start by assuming that the boundary supports phase fluctuations of a local U$(1)$ order parameter, $\Phi$, with charge $Q$, 
\begin{align}
\Phi\sim e^{iQ\sigma(\rho)}\,, 
\end{align}
where $\sigma$ is a compact scalar degree of freedom. Under U$(1)$, 
\begin{align}
\Phi\rightarrow e^{iQ\alpha}\Phi\,,\qquad\sigma\rightarrow\sigma+\alpha\,.
\end{align}
We keep the charge $Q$ general to accommodate a range of possible situations: For example, if $\Phi$ is a superconducting order parameter, it should have charge $Q=2$. 

The coupling of the boundary phase fluctuations to the bulk degrees of freedom must be consistent with the Ward identity and its corollaries, Eqs.~\eqref{eq: boundary Ward general} -- \eqref{eq: tilt total derivative}, along with the periodicity of $\sigma$. These requirements fix the leading terms in the boundary effective action,
\begin{align}
\label{eq: boundary action}
S_{\mathrm{boundary}}=\int d^2\rho\,\mathcal{L}_{\mathrm{boundary}}=\int d^2\rho\,\left(\frac{1}{2g}(\partial_m\sigma)^2-\sin (Q\sigma)\, J^y+\dots\right)\,,
\end{align}
up to further non-linear terms in $\sigma$ and infinitesimal symmetry breaking fields (which gap $\sigma$ and act as IR regulators). Here $J^y(\rho)$ is the bulk current operator evaluated on the boundary, or equivalently the tilt, and $g$ is a coupling constant. Importantly, we have chosen a normalization where the coupling to $J^y$ is unity~\cite{Metlitski:2020cqy,Padayasi:2021sik,Cuomo:2023qvp}. We will see below why this is necessary for consistency with U$(1)$ invariance. 

First, we observe that the Ward identities developed above are consistent with the equations of motion for $\sigma$, 
\begin{align}
    \frac{\delta\mathcal{L}_{\mathrm{boundary}}}{\delta\sigma(\rho)}=\frac{1}{g}\partial^m\partial_m\sigma+Q\,J^y+\dots=0\,,
\end{align}
where the ellipses again refer to higher-derivative contributions and possible non-linearities, which will not impact our analysis. To linear order in $\sigma$, then, we confirm the status of $J^y$ as a total derivative as in Eq.~\eqref{eq: tilt total derivative}, 
\begin{align}
\label{eq: Jy to sigma}
J^y&=-\frac{1}{Q\,g}\partial^m\partial_m\sigma+\dots
\end{align}

Thus, $\sigma$'s (linearized) equation of motion supplies the boundary constraint necessary for preserving charge conservation. The reason is that local U$(1)$ transformations on the boundary shift ${\sigma(\rho)\rightarrow\sigma(\rho)+\alpha(\rho,y=0)}$, which is the same as varying $\sigma$. Hence, phase fluctuations on the boundary are locked to fluctuations of the current normal to the boundary.

\subsection{Symmetry constraints on boundary OPE coefficients}

We may further leverage global U$(1)$ invariance to constrain operator product expansions (OPEs) of bulk operators as they approach the boundary. Let $\Phi(r)$ be the extension of the order parameter to the bulk. 
If the bulk system is critical but the boundary at $y=0$ is ordered, $\Phi$ should acquire a vacuum expectation value that is uniform in the $x$ -- $\tau$ plane, as symmetry breaking fields are taken to zero. Up to multiplication by a constant phase, there is only one possibility consistent with bulk conformal invariance~\cite{Cardy:2004hm},
\begin{align}
\label{eq: Phi vev}
\langle \Phi(y)\rangle &= \frac{a_\Phi}{(2y)^{\Delta_\Phi}}\,,
\end{align}
meaning that the order parameter decays as a universal power law into the bulk. Here $\Delta_\Phi$ is the bulk scaling dimension of $\Phi$ (without coupling to a boundary), and $a_\Phi$ is a real, universal constant.

We wish to establish the relationship between the fluctuations of $\Phi$ and the boundary phase fluctuations, $\sigma$. Global U$(1)$ invariance places tight constraints on correlation functions of $\Phi$ with boundary operators. Say we act with a global U$(1)$ transformation. Then the vacuum expectation value, Eq.~\eqref{eq: Phi vev}, should transform as ${\langle\Phi(y)\rangle\rightarrow e^{iQ\alpha}\,\langle\Phi(y)\rangle}$. This symmetry action must be reproduced self-consistently on taking $\sigma\rightarrow\sigma+\alpha$ in the path integral. To linear order in $\alpha$, this requirement amounts to a Ward identity~\cite{Metlitski:2020cqy}, 
\begin{equation}
   i \langle \Phi(y)\rangle=\int d^{2}\rho' \;\langle \Phi(\rho,y) J^{y}(\rho')\rangle\,,
    \label{eq: second boundary ward identity}
\end{equation}
where the integral is taken over the boundary. Descendent Ward identities for higher-point functions can be obtained by imposing U$(1)$ invariance self-consistently to higher and higher orders in $\alpha$. 
Equation~\eqref{eq: second boundary ward identity} indicates that correlations between the order parameter in the bulk and the current fluctuations on the boundary cannot vanish. We are therefore led to define the boundary OPE,  
\begin{equation}
    \Phi(\tau,x,y\rightarrow 0)\sim \langle \Phi(y)\rangle+iC_{\Phi J}\,(2y)^{2-\Delta_{\Phi}}J^{y}(\rho)+\dots\,,
    \label{eq: boundary OPE 1}
\end{equation}
where the ellipses denote less singular terms and $C_{\Phi J}$ is a real, universal OPE coefficient defined up to a phase. The boundary OPE may be thought of as a traditional OPE between a bulk operator and a defect operator implementing boundary conditions. 
Equation~\eqref{eq: boundary OPE 1} manifests the connection between the bulk and boundary variables, as $J^y(\rho)$ is related to the boundary phase fluctuations through Eq.~\eqref{eq: Jy to sigma}. 

The boundary OPE coefficient, $C_{\Phi J}$, satisfies an exact relation with the boundary current-current correlator,  
\begin{align}
    \langle J^y(\rho)J^y(0)\rangle &=\frac{C_{JJ}}{|\rho|^{4}}\,.
    \label{eq: cor of j 1}
\end{align}
Inserting the definition of the boundary OPE, along with the vacuum expectation value~\eqref{eq: Phi vev}, into the Ward identity, Eq.~(\ref{eq: second boundary ward identity}), we obtain~\cite{Bray:1977fvl,Metlitski:2020cqy},
\begin{align}
    C_{\Phi J}=\frac{a_{\Phi}}{4\pi}\frac{1}{C_{JJ}}\,.
\end{align}
We emphasize that although the OPE coefficients appearing in this equation are universal, real numbers, both $a_\Phi$ and $C_{\Phi J}$ are only defined up to a constant phase, i.e. both sides of the equation should be understood as transforming under U$(1)$.

\subsection{RG flow to the extraordinary fixed point}
\label{sec. extraordinary log RG}

Having established the boundary effective action in Eq.~\eqref{eq: boundary action}, one can compute the perturbative renormalization group (RG) flow of the coupling, $g$, following Ref.~\cite{Metlitski:2020cqy}. We continue to focus on the example U$(1)$ symmetry, where $g$ is exactly marginal in the absence of any bulk degrees of freedom~\cite{Brezin:1976qa}.

Integrating out bulk fluctuations leads to an effective boundary action,
\begin{align}
   S_{\mathrm{eff}}=&\int d^2\rho\, \frac{1}{2g}(\partial_{m}\sigma(\rho))^2-\frac{Q^2}{2}\int d^2\rho \,d^2\rho'\,\sigma(\rho)\, G_{J^yJ^y}(\rho-\rho')\,\sigma(\rho')+\mathcal{O}(\sigma^4)\,,
\end{align}
where ${G_{J^yJ^y}(\rho-\rho')=\langle J^y(\rho)J^y(\rho')\rangle}$ is the current-current correlation function. 

Because $G_{J^yJ^y}$ is nonvanishing at zero boundary momentum, a constant counterterm must be introduced to preserve the U$(1)$ shift symmetry of $\sigma$. Including this term and computing $G_{J^yJ^y}(p)$ to order $p^2$ reveals a logarithmic divergence,
\begin{align}
    G_{J^yJ^y}(p)-G_{J^yJ^y}(0)= \int d^2\rho\, \frac{e^{ip\cdot \rho}-1}{|\rho-\rho'|^{4}}=-\frac{\pi}{2}\,C_{JJ}\,p^2\log \frac{\Lambda}{p}\,,
\end{align}
where we have introduced a hard UV cutoff, $\Lambda$. This computation indicates that $g$ decreases under the RG. Introducing a reference energy scale, $\mu$, an associated differential RG length, ${d\ell\sim -d\log \mu}$, and a running coupling, ${\overline{g}=Z_g^{-1} g}$, ${Z_{g}=(1+\frac{\pi}{2} Q^2C_{JJ}d\ell)}$, leads to an RG flow,
\begin{align}
    \frac{d\overline{g}}{d\ell}=-\frac{\pi}{2}\,Q^2\,C_{JJ}\,\overline{g}^2,
\end{align}
Hence, if $C_{JJ}>0$, the coupling, $g$, indeed runs to zero. Nevertheless, the effect of the bulk-boundary coupling at this fixed point is not innocuous. 

Let $\varphi\sim e^{iQ\sigma}$ be the boundary order parameter. Including the order parameter anomalous dimension and integrating the Callan-Symanzik equation implies that the boundary correlations do not decay as a power law, as in typical examples of quasi-long-ranged order. Instead, they decay much more weakly, as a universal power of $\log(\mu\rho)$ in the limit of large boundary separation~\cite{Metlitski:2020cqy}, 
\begin{align}
\langle\varphi^\dagger (\rho)\varphi(0)\rangle\sim \frac{1}{[\log(\mu\rho)]^{q}}\,,\qquad\rho\rightarrow\infty\,,
\end{align}
with a universal exponent,
\begin{align}
\label{eq: q general}
    q=\frac{1}{\pi^2} \frac{1}{C_{JJ}}\,.
\end{align}
Boundaries such as these are said to possess \emph{extraordinary-log} correlations. The original work of Ref.~\cite{Metlitski:2020cqy} found that such fixed points are possible in O$(n)$ models with ${2\leq n < n_{\mathrm{crit}}\approx 4}$. Models with U$(1)$ symmetry correspond to the case ${n=2}$, for which Ref.~\cite{Metlitski:2020cqy} finds ${C_{JJ}=1/2\pi^2}$, ${q=2}$. Although similar logarithmic corrections to scaling occur in other situations (e.g. $|\phi|^4$ theory in $D=4$ dimensions) involving marginally irrelevant couplings, the extraordinary fixed point is relatively unique in that these corrections supply the primary spatial dependence of the order parameter correlations. 

We now turn to the task of extending this understanding beyond O$(n)$ models, to the particular case of the NCCP$^{N-1}$ model with fermion edge modes. 

\section{The boundary of the SC -- QSH transition}
\label{sec: extra-ordinary NCCP$^{N-1}$ model}

Starting from the conservation laws of a critical system with a boundary, we have seen that it is possible to obtain the extraordinary-log exponent, $q$, simply by knowing the residue of the current-current correlator normal to the boundary, $C_{JJ}$. Our goal in this work is to leverage this simple fact to study the edge of a large-$N$ generalization of the QSH -- SC transition. We focus on a choice of boundary conditions where the helical boundary fermions inherited from the QSH phase couple strongly to the gapless bulk, and we will see that this situation coincides with one where boundary Cooper pair operators -- which in the bulk field theory correspond to monopoles -- exhibit quasi-long-ranged order with extraordinary-log correlations. In this Section, we introduce the NCCP$^{N-1}$ model describing the transition alongside its boundary effective action, and we explain our choice of boundary conditions. The discussion here will set up the ultimate large-$N$ calculation of extraordinary-log correlations in Section~\ref{sec: calculation}. 

\subsection{The QSH -- SC DQCP, with an edge}

We start by reviewing some of the basic aspects of the QSH -- SC DQCP~\cite{Grover2008}. The microscopic degrees of freedom consist of spinful electrons with SU$(2)$ spin rotation and U$(1)$ charge conservation symmetries. When the spin rotation symmetry is broken spontaneously down to U$(1)_z$, corresponding to $S_z$-conservation, the electrons find themselves in a QSH insulating state with helical boundary modes. A direct transition to a $s$-wave superconducting state restoring the full SU$(2)$ symmetry is possible on condensing unit skyrmions, which in such a system carry charge-2$e$ due to the quantum spin Hall response, leading to superconductivity.

The bulk field theory describing this transition is the non-compact $\mathbb{CP}^1$ (NCCP$^1$) model, which consists of two complex scalar boson species, $\phi_I$, $I=1,2$, coupled minimally to an emergent U$(1)$ gauge field, $a_\mu$,
\begin{align}
\label{eq: NCCP1}
S_{\mathrm{bulk}}=\int d^3r\,\left[ (D_{\mu}\phi_I)^{\dagger}D_{\mu}\phi_I+\frac{u}{4}\left(|\phi_1|^2+|\phi_2|^2\right)^2+\dots\right]\,,
\end{align}
where $D_{\mu}=\partial_\mu-ia_\mu$, $u$ is a coupling constant, and repeated flavor indices are summed over. The ellipses denote operators which are irrelevant in the renormalization group sense, such as the Maxwell term for $a_\mu$ (which will be resuscitated later for the large-$N$ limit).

The U$(1)$ gauge theory is non-compact\footnote{Here we adopt the standard condensed matter usage. Importantly, the gauge theory we consider is compact in the sense common to the high energy literature: The gauge group is U$(1)$, not $\mathbb{R}$, meaning that monopole operators do exist even if they are absent from the action.}, in the sense that monopole operators, $\mathcal{M}_a$, are available, but they are not included in the action itself. Consequently, we may discuss the global conservation of monopole number,
\begin{align}
J^\mu_{\mathrm{top}}=\frac{i}{2\pi}\varepsilon^{\mu\nu\lambda}\partial_\nu a_\lambda\,,\qquad \partial_\mu J^\mu_{\mathrm{top}}=0\,.
\end{align}
This symmetry is interpreted as EM charge conservation. The unit monopoles in this theory are bosons, so they carry charge-$2e$ and transform under the associated U$(1)_{\mathrm{top}}$ symmetry as ${\mathcal{M}_a\rightarrow e^{2i\alpha}\mathcal{M}_a}$. They correspond to the charged skyrmions mentioned above, although one may also simply think of them as Cooper pairs. In addition, the theory is invariant under a global SU$(2)$ spin rotation symmetry, which acts as $\phi_I\rightarrow U_{IJ}\phi_{J}$, with $U_{IJ}$ a unitary matrix.

Meanwhile, to capture the correct boundary physics of the QSH phase, it is necessary to introduce helical fermion fields, $\psi_{p}$, ${p=L,R}$, carrying charge under $a_\mu$,
\begin{align}
\label{eq: Fermi edge}
S_{\mathrm{boundary}}=\int d^2\rho\left[\psi^\dagger_p(\partial_\tau+ia_\tau)\psi_p-i\psi^\dagger_{p}\,\sigma^z_{pq}(\partial_x+ia_x)\psi_q\right]+S_{\mathrm{int}}\,.
\end{align}
Here $S_{\mathrm{int}}$ contains the bulk-boundary interactions, which depend on the choice of boundary conditions for bulk fields, so we table their discussion for now. Although the vector U$(1)$ symmetry, ${\psi\rightarrow e^{i\alpha}\psi}$, is gauged, the fermions carry unit charge under the axial U$(1)$ symmetry, ${\psi\rightarrow e^{i\alpha\sigma^z}\psi}$, which is identified\footnote{The identification of axial rotations with the global EM charge conservation symmetry, U$(1)_{\mathrm{top}}$, is a choice. Without affecting any of our results, one may also choose edge variables such that U$(1)_{\mathrm{top}}$ acts with the same charge on the two fermion helicities, in which case $a_\mu$ couples to their axial current.} with the action of U$(1)_{\mathrm{top}}$. 
Although the axial symmetry would be broken anomalously in a purely 1$d$ system, it is preserved here through inflow with the higher-dimensional bulk, which may be regarded as an intrinsically gapless SPT~\cite{Thorngren2021,Potter2023}. Indeed, the microscopic necessity of these boundary fermions and the  nullification of their anomaly at the DQCP can be motivated using a parton construction~\cite{Ma:2021dfx,myerson2024pristine}, although for us it will suffice to check that they support the desired bulk phase diagram. 

The phase diagram may be explored by introducing a $\mathrm{SU}(2)$-invariant mass, $r|\phi_I|^2$. 
For $r<0$, the matter bosons condense, $a_\mu$ is Higgsed, and the $\mathrm{SU}(2)$ symmetry is broken spontaneously down to U$(1)_z$, with the U$(1)_{\mathrm{top}}$ charge conservation symmetry also remaining. This is the QSH phase: The edge modes in Eq.~\eqref{eq: Fermi edge} are liberated of their coupling to the bulk gauge field, and their vector U$(1)$ symmetry is promoted to the global U$(1)_z$ symmetry. This phase also possesses one remaining Goldstone mode.

On the other hand, for $r>0$, the $\phi_I$ fluctuations are gapped, leaving the remaining gauge fluctuations to generate a finite monopole vacuum expectation value, $\langle\mathcal{M}_a\rangle\neq 0$. The gauge theory is then said to be in a Coulomb phase and hosts a gapless photon. Because the monopoles are charge-$2e$ Cooper pairs, their condensation spontaneously breaks U$(1)_{\mathrm{top}}$ down to $\mathbb{Z}_2$, and the resulting Coulomb phase is a superconductor. In turn, the edge modes acutely feel the photon fluctuations, in the form of a logarithmic confining potential that forces them to assemble into Cooper pairs, $\varepsilon_{pq}\psi^\dagger_p\psi_q$, which are neutral under $a_\mu$ but carry EM charge-$2e$. These Cooper pairs are indistinguishable from the bulk superconducting condensate, so we conclude that the edge modes are completely erased, as necessary for a $s$-wave superconductor. 

The NCCP$^1$ model in Eq.~\eqref{eq: NCCP1} has also been proposed to describe the transition between the N\'{e}el and VBS phases of the spin-1/2 square lattice Heisenberg antiferromagnet~\cite{senthil2004deconfined} (see Ref.~\cite{Senthil:2023vqd} for a review of recent developments), with the main distinguishing feature of the QSH -- SC transition being the existence of microscopic fermions. We remark that in both cases, the model has famously been conjectured to describe a continuous transition enjoying an emergent SO$(5)$ symmetry~\cite{Nahum2015,Wang:2017txt}. However, numerical calculations using classical Monte Carlo~\cite{Chen:2013ap,Nahum:2015jya}, conformal bootstrap~\cite{Nakayama:2016jhq,Poland:2018epd}, and fuzzy sphere~\cite{Zhou2024} methods suggest a scenario where the theory flows to a complex, non-unitary fixed point~\cite{Wang:2017txt,Gorbenko:2018ncu,Gorbenko:2018second,Nahum:2019fjw,Ma:2019ysf}, meaning that it would instead describe a fluctuation-induced first-order transition with very large but finite correlation length. Another possibility being explored is that a continuous transition exists but is in fact multicritical~\cite{Takahashi:2024xxd}. In each of these cases, the correlation length is expected to be so large that any first-order behavior would be almost invisible to local correlation functions at finite temperature, meaning that universal features deduced assuming a continuous transition may remain physically relevant. 

\subsection{Boundary conditions and gauge fixing}

Our goal is to study the boundary dynamics of the NCCP$^{N-1}$ model in the $N\rightarrow\infty$ limit, extrapolating our results to finite values of $N$. Introducing a boundary at $y=0$, we will adopt a boundary condition preserving SU$(N)$ symmetry, where the matter fields are turned off. Such a boundary condition implies that gauge fluctuations become stronger as they near the boundary, where screening due to the gapless matter in the bulk becomes weaker. In turn, the monopoles -- which we recall are interpreted microscopically as Cooper pairs -- tend toward condensation as the boundary is approached. The consequence is that these partially screened gauge fluctuations will mediate extraordinary-log correlations among Cooper pairs on the boundary. We will comment on the alternate choice of boundary conditions -- which breaks SU$(N)$ and is less amenable to a large-$N$ limit -- at the end of Section~\ref{sec: calculation}. A further set of boundary conditions was considered in Ref.~\cite{Ma:2021dfx}, leading to different results.

More precisely, we assume the monopoles acquire a vacuum expectation value, 
\begin{align}
\label{eq: monopole vev}
\langle\mathcal{M}_a(y)\rangle=\frac{a_{\mathcal{M}_a}}{(2y)^{\Delta_{\mathcal{M}_a}}}\,,
\end{align}
corresponding to quasi-long-ranged order of the the U$(1)_{\mathrm{top}}$ symmetry at the boundary. 
The gauge invariant boundary condition implementing Eq.~\eqref{eq: monopole vev} is
\begin{align}
\label{eq: BC}
\phi(y=0)=0\,,\qquad e_y(y=0)=0\,,\qquad \partial_ye_x(y=0)=0,\qquad b(y=0)=0\,,
\end{align}
where $e_x=if_{x\tau}$, $e_y=if_{y\tau}$, $b=f_{xy}$ are respectively the emergent electric field normal to the boundary and the (scalar) emergent magnetic field. 

To translate Eq.~\eqref{eq: BC} into  boundary conditions on individual components of the gauge field, $a_\mu$, we must choose a gauge. A natural choice is to keep only the gauge fluctuations with momentum transverse to the boundary, 
\begin{align}
\label{eq: gauge fixing condition}
\partial_x a_x+\partial_\tau a_\tau=0\,.
\end{align}
Taken together with Eq.~\eqref{eq: BC}, this gauge choice uniquely fixes
\begin{align}
\label{eq: a BC}
a_y(y=0)=0\,,\qquad \partial_y a_x(y=0)=\partial_y a_\tau(y=0)=0\,,
\end{align}
leaving the electric field along the boundary, $e_x$, as the only remaining degree of freedom.

At the boundary defined by Eq.~\eqref{eq: BC}, fluctuations of the matter variables, $\phi$, become  gapped. What remains is the component of the electric field \emph{along} the boundary, $e_x$, which is proportional to the monopole current \emph{normal} to the boundary, $J^y_{\mathrm{top}}$. This means that monopoles are allowed to pass through the boundary, so long as the monopole density -- the emergent magnetic field, $b(y=0)/2\pi$ --  vanishes, thereby preserving the monopole conservation symmetry. Because the boundary OPE of the order parameter, $\mathcal{M}_a$, contains the normal component of the current, $J^y_{\mathrm{top}}$, these conditions are consistent with our desired ordered boundary.

\subsection{Boundary effective theory}

Given these boundary conditions and the discussion in Section~\ref{sec. boundary ward identity}, we determine the coupling of the boundary fermions to the bulk critical fluctuations. This may be naturally achieved using bosonization. We introduce two compact scalars, $\sigma$ and $\vartheta$, on the boundary, such that
\begin{align}
\psi_L\sim e^{i(\vartheta-\sigma)}\,,\qquad \psi_R\sim e^{i(\vartheta+\sigma)}\,.
\end{align}
Here the dual field, $\vartheta$, carries charge under the emergent gauge field, $a_\mu$, while $\sigma$ is gauge invariant and carries unit charge under U$(1)_{\mathrm{top}}$. The superconducting order parameter on the edge may therefore be assembled as
\begin{align}
\Delta_{\mathrm{SC}}=\varepsilon_{pq}\psi^\dagger_p\psi_q\sim e^{2i\sigma(\rho)}\,.
\end{align}
Under bosonization we may therefore rewrite Eq.~\eqref{eq: Fermi edge} in terms of $\sigma$ and the remaining boundary gauge components, $a_{m}$, 
\begin{align}
S_{\mathrm{boundary}}=\int d^2\rho\left(\, \frac{1}{2g}(\partial_m\sigma)^2+\frac{i}{\pi}\,\varepsilon_{mn}a_m\partial_n\sigma+\dots\,\right)\,,
\end{align}
where the ellipses denote higher-derivative and non-linear terms in $\sigma$. Integrating by parts allows us to rewrite the action to more explicitly resemble the linearized form of Eq.~\eqref{eq: boundary action}, with $J^y_{\mathrm{top}}=-if_{x \tau}/2\pi$, 
\begin{align}
    S_{\mathrm{boundary}}=\int d^2\rho\left(\, \frac{1}{2g}(\partial_m\sigma)^2+i(2\sigma)\,\frac{f_{x\tau}}{2\pi}+\dots\right)\,.
    \label{eq: ah effective boundary action}
\end{align}

Crucially, the electric field, $f_{x\tau}$, has support in the bulk as well as on the boundary. In contrast, $\sigma$ resides only on the boundary. Notice that this is the same action we would have obtained had we introduced couplings to monopole operators of the form ${ie^{-2i\sigma}\mathcal{M}_a+\mathrm{h.c.}}$, expanded $\mathcal{M}_a$ using the boundary OPE in Eq.~\eqref{eq: boundary OPE 1}, and linearized the result in $\sigma$.  

We will study the combined critical bulk plus boundary model, $S_{\mathrm{bulk}}+S_{\mathrm{boundary}}$, in the limit where the system preserves the U$(1)_{\mathrm{top}}$ symmetry. Consequently, we require that sources for any monopole operators in the Lagrangian, $h(x,y,\tau)\mathcal{M}_{a}$, vanish for all $y>0$. If monopole operators are turned on at $y=0$, U$(1)_{\mathrm{top}}$ will be broken explicitly, and $\sigma$ will be gapped out. This behavior is analogous to the behavior of the O$(2)$ model in the presence of a symmetry-breaking boundary. In the calculations below, we will work in the limit $h(y=0)\rightarrow 0$.

\section{Extraordinary-log exponent in the large-$N$ limit}
\label{sec: calculation}

\subsection{Strategy}

Our interest will be in correlation functions of the boundary SC order parameter, ${\Delta_{\mathrm{SC}}=e^{2i\sigma}}$. As described in Section~\ref{sec. boundary ward identity}, this operator is anticipated to exhibit extraordinary-log correlations, with exponent $q$ determined by the correlator of $J^y_{\mathrm{top}}=-if_{x\tau}/2\pi$ as in Eq.~\eqref{eq: q general}. Since the current is a derivative of the gauge field, the current-current correlator can be extracted by differentiating the gauge propagator. To facilitate this calculation, we consider the problem with $N$ species of bulk matter fields, known as the NCCP$^{N-1}$ model, and we perform an expansion in powers of $1/N$. Using this expansion, our task is to compute the gauge propagator self-consistently in the large-$N$ limit by solving the NCCP$^{N-1}$ model's Schwinger-Dyson equations.

\subsection{The NCCP$^{N-1}$ model}

The large-$N$ generalization of the QSH -- SC transition is the NCCP$^{N-1}$ model, consisting of $N$ complex scalar boson species, $\phi_I$, ${I=1,\dots,N}$, coupled to a fluctuating U$(1)$ gauge field, $a_\mu$, in $D=3$ Euclidean spacetime dimensions, 
\begin{align}
   S_{\mathrm{bulk}}=&\int d^3r\,\left[ \frac{N}{4 e^2}f_{\mu\nu}^2+(D_{\mu}\phi_I)^{\dagger}D_{\mu}\phi_I+\frac{N}{2u}\lambda^2+i\lambda\, \phi_I^{\dagger}\phi_I\right],\; I=1,...,N\,.
   \label{eq: AH action 1}
\end{align}
Here ${f_{\mu\nu}=\partial_\mu a_\nu-\partial_\nu a_\mu}$, ${\mu,\nu=x,y,\tau}$, is the Euclidean field strength; $e$ is the gauge coupling constant. We have introduced the Hubbard-Stratonovich field, $\lambda$, with equation of motion 
\begin{align}
i\lambda= \frac{u}{N}|\phi_I|^2\,, 
\end{align}
to decouple the scalar self-interactions. We take the theory to be at criticality with vanishing bare mass for $\phi_I$. In addition to $U(1)_{\mathrm{top}}$, the theory displays a SU$(N)$ global symmetry rotating the $\phi_I$'s, which for $N=2$ may be interpreted as spin rotation symmetry. The total global symmetry manifest at the level of the action\footnote{The true flavor symmetry is PSU$(N)$=SU$(N)/\mathbb{Z}_N$, since transformations in the center of the group can always be undone with a gauge transformation.} is therefore ${\mathrm{SU}(N)\times \mathrm{U}(1)_{\mathrm{top}}}$, along with time-reversal symmetry.

Importantly, the deformation to $N$ species of bosons does not affect the boundary fermion fields, which carry charge under U$(1)_{\mathrm{top}}$ but not SU$(N)$ at the critical point. Hence, the edge continues to be populated by a single pair of helical fermions and is identical to Eq.~\eqref{eq: Fermi edge}. A bosonized description in terms of a compact scalar field, $\sigma$ such that $\Delta_{\mathrm{SC}}\sim e^{2i\sigma}$ is again possible and leads to the action\footnote{Strictly speaking, we remark that the boundary charge-1 fermions are only necessary for even values of $N$, with the bulk emergent 't Hooft anomaly vanishing trivially for odd $N$. In those cases, the elementary monopole may carry charge-$1$ under U$(1)_{\mathrm{top}}$, and our boundary order parameter fluctuations may be expressed as $e^{i\sigma}$ instead of $e^{2i\sigma}$. This difference does not quantitatively affect our results.} in Eq.~\eqref{eq: ah effective boundary action}. 

The phase diagram of this model is very similar to its $N=2$ counterpart. When $\phi_I$ condenses, the $\mathrm{SU}(N)$ symmetry is broken down to U$(N-1)$, Higgsing $a_\mu$, and leaving a Goldstone phase with an insulating bulk and helical fermions persisting on the boundary. When $\phi_I$ is gapped, the monopoles, $\mathcal{M}_a$, again condense, breaking U$(1)_{\mathrm{top}}$ and producing a SU$(N)$-singlet superconducting phase.

Note that in addition to facilitating an analytically tractable limit as $N\rightarrow\infty$, the extension to even moderate values of $N>2$ makes the possibility of a continuous transition more reasonable, despite old calculations using the epsilon expansion found that $N>183$ is necessary~\cite{Halperin:1973jh}. Indeed,  numerical calculations have suggested that the critical value of $N$ where a continuous transition is possible lies somewhere between $3$ and $10$~\cite{Lou2009,Harada2013,Bonati:2020jlm}. Recent conformal bootstrap calculations have even suggested that the theory is conformal all the way down to  ${N=3}$~\cite{Chester:2025uxb}. Nevertheless, the $N\rightarrow\infty$ limit is well under control and is known to flow to a conformal fixed point~\cite{PhysRevD.24.2169}.

\subsection{Large-$N$ limit and saddle point equations}

We study the NCCP$^{N-1}$ model with boundary, defined in Eqs.~\eqref{eq: AH action 1} and \eqref{eq: ah effective boundary action}, with boundary conditions in Eq.~\eqref{eq: a BC}. Because the critical bulk theory is strongly coupled, we work in the limit of $N\rightarrow\infty$ matter species, holding the fine structure constant, $e^2$, and the quartic self-interaction strength, $u$, fixed. In the absence of a boundary, the large-$N$ limit corresponds to a well studied conformal field theory~\cite{Halperin:1973jh,PhysRevD.24.2169,Appelquist:1988sr} (for a review, see Ref.~\cite{Moshe:2003xn}). We will adapt this limit to the case with a boundary exhibiting U$(1)_{\mathrm{top}}$ quasi-long-ranged order. 

Before constructing the large-$N$ saddle point equations, we first describe the types of classical solutions for $\phi_I$ and $a_\mu$ we are interested in. First, because the matter variable, $\phi_I$, has Dirichlet boundary conditions, we should seek classical solutions where $\phi$ vanishes everywhere,
\begin{equation}
    \langle\phi_{I}\rangle= 0\,.
\end{equation}
On the other hand, the classical solution for  the monopole operator, $\langle\mathcal{M}_a\rangle$, should have scaling behavior with $y$ in the bulk, consistent with an ordered boundary, as in Eq.~\eqref{eq: monopole vev}.

Although the classical solution for the monopole operator is non-vanishing, this does not result in a non-trivial profile for $\langle f_{\mu\nu}\rangle$, as this would violate monopole conservation, U$(1)_{\mathrm{top}}$. Hence, on fixing the gauge as in Eq.~\eqref{eq: gauge fixing condition}, we may seek solutions with
\begin{equation}
    \langle a_\mu\rangle =0\,.
\end{equation}
Indeed, even though monopoles experience quasi-long-ranged order on the boundary, we may solve self-consistently for correlation functions of bulk fields, e.g. $\phi_I$ and $f_{\mu\nu}$, in the absence of any flux background. Put differently, implementing the nonvanishing monopole expectation value, Eq.~\eqref{eq: monopole vev}, in the bulk is tantamount to implementing the boundary conditions in Eq.~\eqref{eq: BC}. We compute bulk correlation functions respecting these boundary conditions self-consistently in the large-$N$ limit.

\subsubsection{Matter propagator}

We take the limit of $N\rightarrow\infty$ bulk $\phi_I$ species, following the approach of Refs.~\cite{Bray:1977tk,McAvity:1995zd}. The saddle point propagator of $\phi_I$, denoted $\langle \phi^\dagger_I(r)\phi_J(r')\rangle=G_\phi(r,r')\delta^{IJ}$, is determined by 
\begin{align}
    (\partial^{\mu}\partial_{\mu}-i\lambda_c )\,G_{\phi}(r,r')=\delta(r-r')\,.
    \label{eq: correlation function of phi at saddle point}
\end{align}
Here we have gauge fixed as described above and set $\langle a_{\mu}\rangle=0$. Notably, however, we have kept the dependence on the saddle point value of the Hubbard-Stratonovich field, $\langle\lambda\rangle\equiv \lambda_c$, which we will see is non-zero even in the $N\rightarrow\infty$ limit.

We now turn to the saddle point equations governing the correlation functions of $\lambda$. Integrating out $\phi$ and expanding the resulting effective action in powers of $1/N$, we find that the saddle point equation for ${\chi_{\lambda}(r,r')\equiv N(\langle\lambda(r)\lambda(r')\rangle-\langle\lambda(r)\rangle\langle\lambda(r')\rangle})$ is solved by computing the usual bubble diagram, 
\begin{align}
\int d^3 w\left[G_\phi(r,w)\right]^2 \chi_\lambda(w,r')+\mathcal{O}(N^{-1})=\delta(r-r')\,,
\label{eq: lambda SEQ}
\end{align}
where $w_\mu=(w_\tau,w_x,w_y)$ is a bulk spacetime coordinate. We remark that we have taken the infrared limit, $u\rightarrow\infty$, prior to the $N\rightarrow\infty$ limit in obtaining Eq.~\eqref{eq: lambda SEQ}, allowing us to neglect the $N\lambda^2/u$ term in the Lagrangian. 
Note also that in the $N\rightarrow\infty$ limit the saddle point equations equations for $\chi_\lambda$ may be solved independently from those for the gauge propagator, since at one loop the fluctuations of $\lambda$ do not mix with those of $a_\mu$.

 Equations~\eqref{eq: correlation function of phi at saddle point} -- \eqref{eq: lambda SEQ} can be solved self-consistently in the large-$N$ limit. However, compatibility  with the Dirichlet boundary condition, $\phi(y=0)=0$, requires that $\lambda_c$ be non-vanishing~\cite{Bray:1977tk,McAvity:1995zd}. One can motivate its form through a simple scaling argument. The reduced conformal symmetry in the presence of the boundary suggests that $\lambda_c$ should decay as a power law in the coordinate normal to the boundary, $\lambda_c(y)\propto y^{-\Delta_\lambda}$. To fix the exponent, we note that deep in the bulk the scaling dimension of $\lambda$ should  coincide with its value in the absence of a boundary, which in the $N\rightarrow\infty$ limit can be read off from Eq.~\eqref{eq: lambda SEQ} to be $\Delta_\lambda=2$. This suggests the ansatz,
\begin{equation}
    i\lambda_c=\frac{C_\lambda}{y^2},
    \label{eq: lambda ansatz}
\end{equation}
where $C_\lambda$ is a constant. Inserting Eq.~\eqref{eq: lambda ansatz} into Eq.~(\ref{eq: correlation function of phi at saddle point}) and using constraints from the reduced conformal invariance, one can solve for $G_{\phi}(r,r')$ as a function of $C_\lambda$, revealing that the value of $C_\lambda$ corresponds uniquely to one's choice of boundary conditions. For the Dirichlet boundary condition of interest to us, $C_\lambda=-1/4$.We provide a pedagogical review of the details of this computation in Appendix~\ref{app: review of o(n) boundary criticality}. 

The necessity of Eq.~\eqref{eq: lambda ansatz} means that even in the $N\rightarrow\infty$ limit the $\phi_I$ propagator experiences corrections from its Gaussian form due to the implementation of boundary conditions. This effect starkly contrasts with the boundary-free case, where despite correlation functions of $\lambda$ receiving a non-perturbative correction in the large-$N$ limit, the propagator of $\phi_I$ persists unaffected. The final result for the saddle point propagator of $\phi$ is naturally expressed in terms of the two relative coordinates,
\begin{equation}
    \scriptr=(\tau-\tau',x-x',y-y'),\qquad \ell=(\tau-\tau',x-x',y+y')\,,
\end{equation}
which are reflected with respect to one another. One obtains 
\begin{equation}
    G_{\phi}(\scriptr,\ell\,)=\frac{1}{4\pi}\sqrt{\frac{1}{\scriptr^2}-\frac{1}{\ell^2}}=\frac{\sqrt{1-v^2}}{4\pi |\scriptr|}\,,\qquad v=\frac{|\scriptr|}{|\ell|}\,.
    \label{eq: Gphi saddle point}
\end{equation}

Indeed, when evaluated on the boundary, $y=y'=0$, meaning that $v=1$ and $G_\phi$ vanishes, consistent with the Dirichlet boundary condition. On the other hand, the limit where the boundary becomes invisible is $v\to0$. In this so-called coincident limit, one  recovers the expected result for the system without a boundary, ${G_{\phi}(v\to0)\rightarrow 1/(4\pi |\scriptr|)}$. 
We emphasize that while this result is reminiscent of the method of images, such an  approach is only valid for Gaussian theories and is precluded by the finite solution for  $\lambda_c$.

\subsubsection{Photon saddle point equations and one-loop polarization tensor}
\label{subsec: saddle point eqs}

{
\begin{figure}[t]
    \begin{center}
    \begin{tikzpicture}[>={[inset=0,length=8,angle'=30,bend]Stealth}, line width=.7pt]
\begin{feynman}
            \vertex (a1);
            \vertex[right= 2 cm of a1](a2);
            \vertex[below=0.12 cm of a1](a3);
            \vertex[below=0.12 cm of a2](a4);
            \diagram* {
            (a1) -- [photon, blue] (a2),
            (a3) -- [photon, blue, edge label'={ $D_{\mu\nu}$}] (a4)
            };
            \node at (2.5,0) {$=$};
\end{feynman}
\begin{feynman}[xshift=3cm]
            \vertex (a1);
            \vertex[right= 2 cm of a1](a2);
            \diagram* {
            (a1) -- [photon, edge label'={ $D^0_{\mu\nu}$}] (a2)
            };
            \node at (2.5,0) {$+$};
\end{feynman}
\begin{feynman}[xshift=6cm]
            \vertex (a1);
            \vertex[right= 1.5 cm of a1](a2);
            \vertex[right= 2.5 cm of a2](a3);
            \vertex[right= 1.5 cm of a3](a4);
            \vertex[below= 0.12 cm of a4](a5);
            \vertex[below= 0.12 cm of a3](a6);
            \diagram* {
            (a1) -- [photon,edge label'={$D^0_{\mu\rho}$}] (a2),
            (a2)   -- [fermion, half left] (a3),
            (a3)   -- [fermion, half left] (a2),
            (a3) -- [photon,blue] (a4),
            (a6) -- [photon,blue, edge label'={$D_{\sigma\nu}$}] (a5),
            };
            \node at (2.75,0) {$\Pi^{\rho\sigma}$};
\end{feynman}
\end{tikzpicture}
\end{center}
    \caption{Diagrammatic representation of the Schwinger-Dyson equations \eqref{eq: original sd equation 1} -- \eqref{eq: original sd equation 2} for the emergent photon propagator. Here $D^{0}_{\mu\nu}$ represents the tree-level propagator, determined by the Maxwell term. The blue double lines denote the fully resummed photon propagator. We solve these equations self-consistently in the large-$N$ limit.}
    \label{fig: sd equation of photon}
\end{figure}
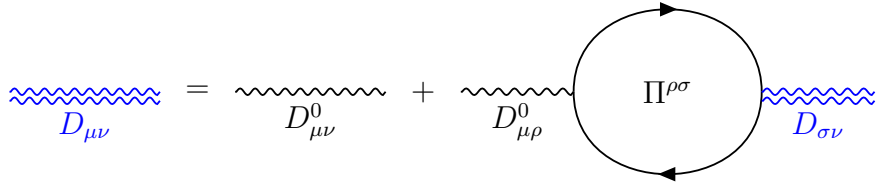
}

We may similarly write down saddle point equations for the photon propagator by integrating out $\phi_I$ and expanding the resulting effective action in fluctuations of $a_\mu$, gauge fixing as in Eq.~\eqref{eq: gauge fixing condition} (for details, see Appendix~\ref{app: extra-ordinary-log ah}). Taking the infrared limit, $e^2\rightarrow\infty$, and varying the effective action produces a pair of formally  exact Schwinger-Dyson equations for the gauge propagator, $D_{\mu\nu}(r,r')=\langle a_\mu(r)a_\nu(r')\rangle$, 
\begin{align}
\label{eq: original sd equation 1}
\int d^3w\,\Pi_{yy}(r,w)D_{yy}(w,r')&=-\delta(r-r')\,.\\
\label{eq: original sd equation 2}
\int d^3w\,\Pi_{ml}(r,w)D_{ln}(w,r')&=-\left(\delta_{mn}-\frac{\partial_{m}\partial_{n}}{|\partial_l|^2}\right)\delta(r-r')\,,  
\end{align}

Here we have defined $\partial'_\mu\equiv \frac{\partial}{\partial r'_\mu}$, and we recall $l,m,n=\tau,x$ are boundary indices. A diagrammatic representation of the above equations is given in Fig.~\ref{fig: sd equation of photon}. 

We may now observe the advantage of the gauge fixing condition in Eq.~\eqref{eq: gauge fixing condition}, $\partial_ma_m=0$. It implies 
\begin{align}
D_{my}=D_{ym}=0\,,
\end{align}
meaning that the Schwinger-Dyson equations separate into independent parallel and normal components, \eqref{eq: original sd equation 1} -- \eqref{eq: original sd equation 2}. This separation follows from the fact that the boundary preserves translation and rotation invariance in the $x$ -- $\tau$ plane. Hence, we may write down a sensible partially Fourier transformed gauge propagator, $\widetilde{D}_{\mu\nu}(k_m,y)$. In particular, covariance under rotations in the $x$ -- $\tau$ plane implies that the only admissible solutions for the mixed components of the propagator have the form, ${\widetilde{D}_{my}=k_mf(k_m,y)}$, where $f$ is a scalar-valued function. However, the gauge fixing condition, Eq.~\eqref{eq: gauge fixing condition}, implies ${k_m\widetilde{D}_{my}=k_mk_m f(k_m,y)=0}$ for all $k_m$, meaning that $f(k_m,y)\equiv0$ and thus $\widetilde{D}_{my}\equiv0$. The same argument may be applied to the transposed components, $\widetilde{D}_{ym}$.

The  polarization tensor, $\Pi_{\mu\nu}(r,r')$, may be computed self-consistently in the $N\rightarrow\infty$ limit as the usual one-loop  diagram, which in real space is
\begin{align}
\Pi_{\mu\nu}(r,r')=2N\left[G_\phi(r,r')\partial_\mu\partial_\nu'G_\phi(r,r')-\partial_\mu G_\phi(r,r')\partial_\nu'G_\phi(r,r')\right]+\mathcal{O}(N^0)\,,
\label{eq: Pi bubble formal}
\end{align}
where $G_\phi(r,r')$ is the saddle point solution in Eq.~\eqref{eq: Gphi saddle point}, and the leading term is $\mathcal{O}(N)$ due to the trace. Gauge invariance of Eq.~\eqref{eq: Pi bubble formal}, ${\partial_\mu\Pi_{\mu\nu}(r,r')=\partial'_\nu\Pi_{\mu\nu}(r,r')=0}$ for $r\neq r'$, follows from the saddle point equation~\eqref{eq: correlation function of phi at saddle point}. Note that formally $\Pi_{\mu\nu}$ also contains a delta function contact term arising for the $a^2|\phi|^2$ seagull vertex, which is necessary for this Ward identity to hold locally for all $r,r'$ (see Appendix~\ref{app: extra-ordinary-log ah}). However, one may choose a short-distance regulator such that this contact term has vanishing contribution to the Schwinger-Dyson equations~\cite{Bray:1977fvl}, so we drop it here. 

We plug the scalar propagator, Eq.~\eqref{eq: Gphi saddle point}, into this expression to obtain the manifestly gauge invariant result, 
\begin{align}
    \Pi_{\mu\nu}(\scriptr,\ell)=&\frac{2N}{(4\pi)^2}\left(\frac{1}{\scriptr^2}-\frac{1}{\ell^2}\right) \left( \frac{\mathcal{T}_{\mu\nu}(\scriptr)}{\scriptr^2}+(1-2\delta_{\nu y})\frac{\mathcal{T}^R_{\mu\nu}(\ell)}{\ell^2}\right),
    \label{eq: polarization tensor}
\end{align}
where summation over repeated indices is not assumed, and $\mathcal{T}_{\mu\nu}(\scriptr)$ and $\mathcal{T}^R_{\mu\nu}(\ell)$ are the generalized transverse projectors, 
\begin{align}
    \mathcal{T}_{\mu\nu}(\scriptr)=\delta_{\mu\nu}-2\frac{\scriptr_{\mu}\scriptr_{\nu}}{\scriptr^2},\qquad
    \mathcal{T}^R_{\mu\nu}(\ell)=\delta_{\mu\nu}-2\frac{\ell_{\mu}\ell_{\nu}}{\ell^2}\,.
\end{align}
Note that because rotation invariance is broken by the presence of the boundary, $\Pi_{m y}\neq \Pi_{ym}$, even though $\Pi_{mn}=\Pi_{nm}$, for boundary indices $n,m=\tau,x$.

\subsection{Extraordinary-log exponent at large-$N$}

\subsubsection{Computation of the photon propagator}

We now turn to the task of computing the photon propagator, which controls the extraordinary-log behavior of the boundary fluctuations. Having obtained the large-$N$ polarization tensor, $\Pi_{\mu\nu}$, in Eq.~\eqref{eq: polarization tensor}, we now plug this result back into the Schwinger-Dyson equations \eqref{eq: original sd equation 1} -- \eqref{eq: original sd equation 2} and solve for the photon propagator. 

Our judicious choice of gauge, Eq.~\eqref{eq: gauge fixing condition}, allows us to  solve for the propagator components in the plane of the boundary, $D_{mn}$, independently from the normal component, $D_{yy}$. We therefore focus on  $D_{mn}$, which is sufficient for computing the correlators of the electric field, ${e_x=if_{x\tau}=-2\pi J^y_{\mathrm{top}}}$, along the boundary and thus the extraordinary-log exponent. 

To attack Eq.~\eqref{eq: original sd equation 1}, we make use of a simplifying trick. For any vector-valued function, $B_m(r,r')$, our gauge choice~\eqref{eq: gauge fixing condition} implies
\begin{align}
0&=\int d^3w \,B_m(r,w)\,\frac{\partial}{\partial w_{l}} D_{ln}(w,r')=-\int d^3w \,\left[\frac{\partial}{\partial w_{l}}B_m(r,w)\right]\, D_{ln}(w,r')\,,
\end{align}
since total derivatives in the plane parallel to the boundary integrate to zero. This means that we may add $\partial B_m/\partial w_l$ to the polarization tensor in Eq.~\eqref{eq: original sd equation 1} without changing the solution for $D_{mn}$,
\begin{align}
\label{eq: trick}
   \int d^3w\,\Pi_{ m l}(r,w)D_{ln}(w,r') &= \int d^3w\,\left[\Pi_{m l}(r,w)-\frac{\partial}{\partial w_{l}} B_{m}(r,w)\right]D_{ln}(w,r')\,.
\end{align}
The freedom to add a total derivative drastically simplifies the problem of solving the integral equation in Eq.~\eqref{eq: original sd equation 1}: Even if the result for the polarization tensor in Eq.~\eqref{eq: polarization tensor} is difficult to invert, we may select a function, $B_{m}(r,r')$, to produce a tensor that can be inverted straightforwardly.

The function we choose has the form, 
\begin{align}
B_m(r,r')&=\frac{N}{8\pi^2}\frac{\partial}{\partial r'_m}\left(\frac{b(v)}{\scriptr^2}\right)\,,\qquad v=\frac{|\scriptr|}{|\ell|}\,.
\end{align}
where
\begin{align}
    b(v)&=\frac{1}{4}\left(v^2-1\right)-\frac{v^2}{(v^2-1)^2}\left(1-v^2+\frac{1+v^2}{2}\log v^2\right)\,.
    \label{eq: subtracting total derivative}
\end{align}
We motivate this choice in Appendix~\ref{app: extra-ordinary-log ah}. In particular, we show that it cancels the off-diagonal components of  $\Pi_{mn}$ to yield 
\begin{align}
    \Pi_{mn}(r,r')-\frac{\partial}{\partial r_n'} B_m(r,r') &=\frac{N}{16\pi^2}\frac{1}{\scriptr^4}\left(\frac{1-v^8+4v^4\log v^2}{(v^2-1)^2}\right)\delta_{mn}\equiv F(r,r')\,\delta_{mn}\,.
\end{align}
Plugging this result back into Eq.~\eqref{eq: original sd equation 1} gives an equation determining the trace of the photon propagator, ${\mathfrak{D}(r,r')\equiv D_{xx}+D_{\tau\tau}}$, 
\begin{align}
    \int  d^3w\, F(r,w)\,\mathfrak{D}(w,r')&=-\delta(r-r')\,.
    \label{eq: FD equation}
\end{align}
This equation can be solved by writing the integrand as a series of hypergeometric functions, following the approach used in Ref.~\cite{McAvity:1995zd}. Leaving the details of the calculation to Appendix~\ref{app: extra-ordinary-log ah}, we find the $N\rightarrow\infty$ solution, 
\begin{align}
    &N\mathfrak{D}=\frac{4}{\pi^2}\frac{v}{\scriptr^2}\left[\frac{2(1-v^2)}{v}
+\pi^2+2 \mathrm{Li}_{2}\left(\frac{(1-v)^2}{(1+v)^2}\right)-8\mathrm{Li}_{2}\left(\frac{1-v}{1+v}\right)
\right]+\mathcal{O}(N^{-1})\,,
\label{eq: trace D result}
\end{align}
where we emphasize the factor of $N$ accompanying $\mathfrak{D}$ on the left-hand side, which allows a nonvanishing result for the right-hand side the strict $N\rightarrow\infty$ limit.

It is straightforward to check that the result in Eq.~\eqref{eq: trace D result} satisfies the necessary consistency conditions. First, in the coincident limit, $v\rightarrow0$ (i.e. $\scriptr\rightarrow 0$ with $\ell\neq 0$), the polylogarithms cancel with the constant term, leaving only 
\begin{align}
   \lim_{\scriptr\rightarrow0} N\mathfrak{D}(\scriptr,\ell\neq0)=\frac{8}{\pi^2}\frac{1}{\scriptr^2}\,,
\end{align}
which matches the Fourier transform of the well known result for the $N\rightarrow\infty$ limit of the NCCP$^{N-1}$ model without boundary~\cite{Moshe:2003xn}. We also find that $\mathfrak{D}$ is consistent with the Neumann boundary conditions in Eq.~\eqref{eq: a BC}, 
\begin{align}
\partial_y \mathfrak{D}(y=0)=\partial'_y\mathfrak{D}(y'=0)=0\,.
\end{align}
Indeed, this boundary condition is not implemented \emph{a priori}. Rather, it arises self-consistently on solving Eq.~\eqref{eq: original sd equation 2} with $\Pi_{mn}$ computed from matter variables respecting the Dirichlet boundary conditions in Eq.~\eqref{eq: BC}. 

In the boundary limit, $y,y'\rightarrow0$, $\scriptr=\ell$, and only the second term in Eq.~\eqref{eq: trace D result} survives, giving in the $N\rightarrow\infty$ limit
\begin{align}
\label{eq: boundary D}
  N \mathfrak{D}(\rho-\rho';y=y'=0)=\frac{4}{|\rho-\rho'|^2}\,,
\end{align}
where we again adopt the notation, $\rho=(\tau-\tau',x-x')$, for the relative coordinate on the boundary. We will see below that this result is sufficient for computing the extraordinary-log exponent on the boundary. 

Nevertheless, before proceeding, we comment on the remaining components of the photon propagator. Although computing them directly by solving Eqs.~\eqref{eq: original sd equation 1} -- \eqref{eq: original sd equation 2} is a challenge we leave to future work, we remarkably find that the correlation functions of the emergent field strength, $f_{\mu\nu}$, can be obtained from knowledge of $\mathfrak{D}$ alone. This result follows from the tight constraints of the Ward identity for $J_{\mathrm{top}}^\mu$, Eq.~\eqref{eq: boundary Ward general}, along with the reduced conformal invariance of the bulk. The solutions indeed respect the boundary conditions in Eq.~\eqref{eq: BC} for the emergent electric and magnetic fields, and they are presented in Appendix~\ref{app: extra-ordinary-log ah}.

\subsubsection{Extraordinary-log correlations}

Equipped with Eq.~\eqref{eq: trace D result}, we are now prepared to compute the correlation functions of the superconducting order parameter at the boundary, ${\Delta_{\mathrm{SC}}\sim e^{2i\sigma}}$, where we recall that the compact scalar field $\sigma$ couples to the Euclidean electric field fluctuations, as in Eq.~\eqref{eq: ah effective boundary action}. Following the discussion in Section~\ref{sec. extraordinary log RG}, we integrate out the bulk degrees of freedom to obtain an effective action, 
\begin{align}
   S_{\mathrm{eff}}=&\int d^2\rho\, \frac{1}{2g}(\partial_{m}\sigma)^2+\frac{2}{(2\pi)^2}\int d^2\rho \,d^2\rho'\,\sigma(\rho)\, \langle f_{x\tau}(\rho)\,f_{x\tau}(\rho') \rangle\,\sigma(\rho')+\mathcal{O}(\sigma^4)\,.
\end{align}
To determine the existence of extraordinary-log correlations, we simply need to evaluate the second term.

Serendipitously, the $f_{x\tau}$ correlator can be expressed entirely in terms of $\mathfrak{D}$. To see this, we exploit some useful properties of the photon propagator on the boundary. First, translation invariance in the plane of the boundary implies ${D_{mn}(\rho,\rho';y=y'=0)=D_{mn}(\rho-\rho')}$ is a function solely of the relative coordinates, $\rho_m-\rho'_m$. Now we make use of the gauge choice in Eq.~\eqref{eq: gauge fixing condition}, which may be recast on the boundary as
\begin{align}
\label{eq: gauge condition propagator}
\partial_m D_{mn}(\rho-\rho')=\partial_n D_{mn}(\rho-\rho')=0\,,
\end{align}
where derivatives are taken with respect to $\rho-\rho'$. Applying this condition alongside the result in Eq.~\eqref{eq: boundary D}, we conclude that on the boundary ($y=y'=0$),
\begin{align}
   \langle J_{\mathrm{top}}^y(\rho) J_{\mathrm{top}}^y(\rho')\rangle &=-\frac{1}{(2\pi)^2}\langle f_{x\tau}(\rho)\,f_{x\tau}(\rho')\rangle\nonumber\\
   &=\frac{1}{(2\pi)^2}\partial_m\partial_m\,\mathfrak{D}(\rho-\rho')\nonumber\\
   &= \frac{1}{N}\left[\frac{4}{\pi^2}+\mathcal{O}(N^{-1})\right]\frac{1}{|\rho-\rho'|^4}\,,
   \label{eq: Jtop correlator final}
\end{align}
where the second line uses Eq.~\eqref{eq: gauge fixing condition}. At infinite $N$, this expression vanishes, reflecting the vanishing of the photon propagator, $D_{\mu\nu}$, in the strict $N\rightarrow\infty$ limit (even though $ND_{\mu\nu}$ remains finite as $N\rightarrow\infty$). We therefore will assume from here that our results can be extrapolated to $N$ large but finite. 

As reviewed in Section~\ref{sec. extraordinary log RG}, a result of the form \eqref{eq: Jtop correlator final} signals running to a boundary fixed point with extraordinary-log correlations~\cite{Metlitski:2020cqy}, 
\begin{align}
    \langle \Delta_{\mathrm{SC}}(\rho)\,\Delta_{\mathrm{SC}}(0)\rangle\sim \langle e^{2i\sigma(\rho)}e^{2i\sigma(0)}\rangle \sim \frac{1}{(\log \mu \rho)^{q}}\,,
\end{align}
where $\mu$ is again a reference energy scale. A perturbative renormalization group calculation finds that $q$ is determined by the residue, $C_{JJ}$, of the boundary current-current correlator, as in Eq.~\eqref{eq: q general}. Our result in Eq.~\eqref{eq: Jtop correlator final} therefore tells us that the extraodinary-log exponent of the NCCP$^{N-1}$ model at large-$N$ is 
\begin{align}
q&=\frac{N}{4}+\mathcal{O}(N^{0})\,.
\label{eq: final ah q}
\end{align}
Equation~\eqref{eq: final ah q} is the main result of this work. It demonstrates the existence of a large new family of extraordinary-log universality classes parameterized by $N$. To our knowledge, these are the first examples of any extraordinary-log universality classes beyond the O$(2)$ and O$(3)$ models: For direct comparison, the theoretical prediction for the O$(2)$ model is $q=2$~\cite{Metlitski:2020cqy}. That $q$ here grows linearly with $N$ should not come as a surprise, as it recalls the fact that the bulk monopole scaling dimensions are proportional to $N$~\cite{Murthy:1989ps,Borokhov:2002ib,Metlitski2008monopoles}. 
We expect the extrapolation to finite $N$ in Eq.~\eqref{eq: final ah q} is controlled so long as the bulk NCCP$^{N-1}$ model remains in its large-$N$ conformal phase. 

\subsubsection{Comments on the boundary phase diagram}

\begin{figure}[t]
    \centering
    \includegraphics[width=0.6\linewidth]{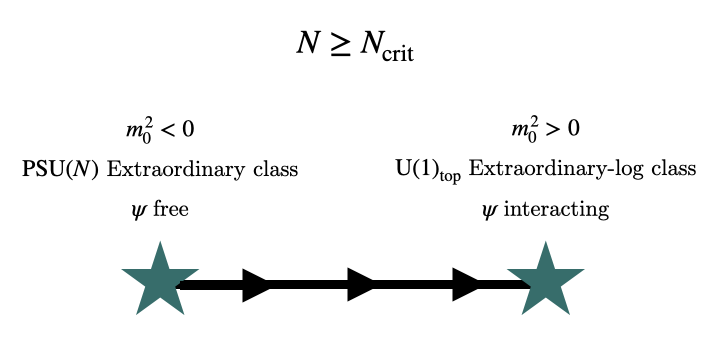}
    \caption{Conjectured large-$N$ renormalization group flow. 
    For $N\geq N_{\mathrm{crit}}$, we propose that the boundary of the deconfined QSH -- SC transition possesses a \emph{unique} boundary phase, where the SC order parameter exhibits extraordinary-log correlations with $q\approx N/4$ and the boundary fermions remain coupled to bulk gauge fluctuations. The corresponding state with $\mathrm{PSU}(N)=\mathrm{SU}(N)/\mathbb{Z}_N$ order at the boundary is expected to be unstable, with its order parameter evolving to an ordinary boundary phase at large-$N$.}
    \label{fig: phase diagram}
\end{figure}

The result in Eq.~\eqref{eq: final ah q} allows us to assemble a proposal for the  boundary phase diagram of the NCCP$^{N-1}$ model. 
Say we implement the boundary conditions on $\phi$ through a mass defect added to the Euclidean action, Eq.~\eqref{eq: AH action 1}\,, 
\begin{align}
\label{eq: mass defect}
S=\int d^3r\,m^2(y)\,|\phi_I|^2+\dots\,,\qquad m^2(y)=m_0^2\,\delta(y)\,.
\end{align}
The problem we have considered up to this point corresponds to the case of $m_0^2>0$, which is the ordinary boundary condition for $\phi$. In the limit $m_0^2\rightarrow\infty$, the boundary fluctuations of $\phi$ are projected out, recovering the Dirichlet boundary condition in Eq.~\eqref{eq: BC} exactly. We demonstrated that this boundary condition is consistent with extraordinary-log boundary correlations for the monopole operators, $\mathcal{M}_a$, which are the order parameters for the U$(1)_{\mathrm{top}}$ charge conservation symmetry and whose condensation in the bulk yields superconductivity. 

In the opposite scenario, $m_0^2<0$, the matter variables will be in their extraordinary boundary condition and tend toward order of the SU$(N)$ (more accurately PSU$(N)$) symmetry near the boundary. 
In the large-$N$ limit, the resulting effective boundary theory should resemble a gauged O$(2N)$ sigma model that interacts with bulk critical fluctuations through the flavor current, ${J^{y,b}_{\mathrm{flavor}}=i\phi^\dagger D_{y} T^b\phi+\mathrm{h.c.}}$, where $T^b$ are the broken generators of SU$(N)$. This sigma model coexists with the gapless boundary fermions, which we anticipate remain nearly free in the large-$N$ limit due to suppression of the emergent electric field near the boundary. We propose that the ultimate fate of this boundary phase depends on the value of $N$.

For $N$ larger than a critical value, $N\geq N_{\mathrm{crit}}$, the large-$N$ limit suggests the boundary sigma model instead flows toward a disordered, gapped phase corresponding to the ordinary boundary condition for the bulk matter. This is none other than the extraordinary fixed point studied in this work, where, in the flow to the IR, gauge fluctuations produce extraordinary-log correlations among the boundary helical fermions. The key takeaway from this observation is that our extraordinary-log  fixed point might be the \emph{unique} symmetry-preserving boundary phase for $N\geq N_{\mathrm{crit}}$, although we cannot rule out the possibility of other phases at intermediate coupling. Our conjectured boundary RG flow is summarized in Fig.~\ref{fig: phase diagram}. 

For ${N<N_{\mathrm{crit}}}$, extrapolation of the large-$N$ theory would suggest that the boundary sigma model flows to an extraordinary fixed point~\cite{Metlitski:2020cqy}, and we expect the boundary spin waves to exhibit extraordinary-log correlations. Between these two extraordinary fixed points, there must exist an ``extra-special'' fixed point at $m_0^2=0$, which was also proposed in Ref.~\cite{myerson2024pristine}. We leave illumination of this fixed point and the estimation of $N_{\mathrm{crit}}$ to future work.

Finally, we remark on the case $N=1$, where the bulk theory is the traditional single-component abelian Higgs model with a U$(1)_{\mathrm{top}}$ monopole conservation symmetry but no global flavor symmetry. This case also corresponds to a possible QSH -- SC transition within the Landau paradigm, where the SU$(2)$ spin rotation symmetry is explicitly broken down to U$(1)_z$ throughout the phase diagram and does not play any role in the transition. When $m_0^2<0$, no spin waves remain on the boundary, which finds itself in a genuinely ordinary phase with no extraordinary-log sector. The helical fermions in this case would remain close to their free fixed point. For $m_0^2>0$, on the other hand, monopoles tend to order at the boundary. The qualitative physics of this phase is essentially identical to what we have explored in this work. However, we caution against extrapolating our results to $N=1$: There is ample evidence that the $N=1$ abelian Higgs model is dual to the XY model in three spacetime dimensions~\cite{Peskin:1977kp,THOMAS1978513,Dasgupta:1981zz}, meaning that the boundary extraordinary-log exponent is expected to be identical to that of the XY model. If the bulk theory becomes weakly first order below a (probably distinct) critical value of $N$, it is unlikely to us that the large-$N$ limit  can be accurately extrapolated to $N=1$ case, where the transition becomes continuous again by duality with the XY model. At any rate, the strong interpretation of duality would predict that the extraordinary-log exponents -- like other universal critical data -- of the XY and abelian Higgs models should match.

\section{Discussion}
\label{sec: discussion}

In this work, we have studied the extraordinary boundary phase of the deconfined quantum spin Hall -- superconductor transition, where quasi-long-ranged SC order on the boundary coincides with bulk criticality. Using a large-$N$ expansion, we computed the extraordinary-log exponent of the boundary superconducting correlations, finding a new family of extraordinary-log universality classes parameterized by $N$. Our large-$N$ calculation can in principle be checked  using Monte Carlo or bootstrap methods, deepening the extent to which the QSH -- SC transition can be a fruitful setting for future numerical study.

Our work constitutes the first explicit construction of extraordinary-log universality classes outside the context of O$(n)$ models, with extraordinary-log behavior being induced on topologically  protected edge degrees of freedom through coupling to the critical bulk. We therefore expect that our approach can be extended to other families of 2$d$ quantum critical systems, such as algebraic spin liquids described by QED$_3$, quantum Hall plateau transitions, or even continuous metal-insulator transitions with spinon Fermi surfaces~\cite{Senthil_2008}. 

We furthermore expect that our work could form a basis to study tunnelling at interfaces between quantum critical systems with superconductors or gapped topological phases, an emerging experimental possibility in the context of 2$d$ materials. For example,  it may be possible to probe exotic boundary correlations  in WTe$_2$ by patterning SC junctions into the material and tuning between QSH and SC phases. 

\section*{Acknowledgements}

We are especially grateful to Max Metlitski for discussions and encouragement at early stages of this work. We thank Zhen Bi, Vladimir Calvera, Shi Chen, Aleksey Cherman, Andrey Chubukov, Eduardo Fradkin, Alex Kamenev, Ruochen Ma, Umang Mehta, Ben Moy, Maria Neuzil, Alex Thomson, and Xiao-Chuan Wu for helpful discussions and comments on the manuscript. The authors are supported by startup funds at the University of Minnesota.

\appendix
\section{Review of boundary criticality in the $\mathrm{O}(n)$ model}
\label{app: review of o(n) boundary criticality}
In this Appendix, we review the boundary phase diagram the $\mathrm{O}(n)$ model in $D=3$ Euclidean spacetime dimensions, with a planar boundary at $y=0$. We follow Refs.~\cite{Bray:1977fvl,McAvity:1995zd,Metlitski:2020cqy,Herzog:2020lel}. 

We start with the bulk action of the O$(n)$ model,
\begin{equation}
    S_{\mathrm{bulk}}=\int d^3r\; \frac{1}{2}(\partial_{\mu}\varphi_I)^2+\frac{u}{2N}((\varphi_I)^2)^2,
\end{equation}
for $I=1,...,n$ species of real boson $\varphi_I$, and the integral is taken over the bulk region $y>0$. We will consider the bulk model at criticality, focusing on the $n\to\infty$ saddle point. Different boundary conditions for $\varphi$ lead to a range of unique boundary scaling behaviors, with universal exponents distinct from their bulk counterparts~\cite{Diehl:1996kd}. Each boundary condition may be conveniently implemented by introducing an O$(n)$-invariant mass term,
\begin{align}
    S_{\mathrm{mass}}=\frac{1}{2}\int d^3r\, \delta(y)\,m_0^2\,(\varphi_I)^2\,.
\end{align}
When ${m_0^2>0}$, $\varphi_I$ will be gapped on the boundary, corresponding to the Dirichlet condition ${\varphi_I(y=0)=0}$. The resulting boundary universality class is called the \emph{ordinary} class. On the other hand, when ${m_0^2=0}$, all components of $\varphi_I$ critically fluctuate at the boundary, corresponding to the Neumann condition ${\partial_y\varphi_I(y=0)=0}$. This yields  the \emph{special} boundary universality class. We remark that the special universality class should be considered an unstable fixed point, as both the bulk and the boundary masses are both tuned to vanish. The final possibility is ${m_0^2<0}$. In this case, $\varphi_I$ condenses on the boundary, breaking $\mathrm{O}(n)$. The resulting system consists of gapless spin waves on the boundary coupled to bulk critical fluctuations, and the combined bulk-boundary system realizes the so-called \emph{extraordinary} boundary  universality class. 

The different boundary universality classes are best exemplified by considering the propagator of $\varphi_I$. The planar boundary at $y=0$ breaks the full conformal group SO$(D,1)$ to the subgroup SO$(D-1,1)$ of transformations preserving the boundary~\cite{Cardy:2004hm}. This reduced conformal group cannot completely constrain the order parameter correlation function to be a single power law, as is usually the case for conformal field theories. Instead, it is necessary to introduce a universal dependence on the crossing ratio, $v$, 
that encodes the absolute distance to the $y=0$ boundary. The general form of the propagator is therefore 
\begin{equation}
    \delta_{IJ}G_{\varphi}(\scriptr,\ell)=\langle \varphi_I(r)\varphi_J(r')\rangle-\langle \varphi_I(r)\rangle\langle\varphi_J(r')\rangle=\frac{\delta_{IJ}}{4\pi|\scriptr|^{2\Delta_\varphi}}F(v),\quad \; v^2=\frac{|\scriptr|^2}{|\ell|^2},
    \label{eq: app prop ansatz}
\end{equation}
where
\begin{equation}
    \scriptr=(\tau-\tau',x-x',y-y'),\qquad \ell=(\tau-\tau',x-x',y+y'),
\end{equation}
For the case of a $D=4$ dimensional bulk, $\Delta_\varphi=D/2-1=1/2$ is simply the Gaussian scaling dimension of $\varphi$ in the $n\to\infty$ limit. Here $F(v)$ is a dimensionless function of the crossing ratio, denoted $v$. In the so-called \emph{coincident limit}, where $\scriptr\to 0$ with $\ell\neq 0$, $v\to 0$, the boundary is essentially projected out, and we should recover the standard result for a system without boundary. Hence, we must fix  
\begin{align}
F(v\rightarrow 0)&=1\,.
\end{align}
For $D\geq4$, the bulk system flows to a Gaussian fixed point. This allows one to compute $F(v)$ using the method of images~\cite{McAvity:1995zd}. For example, the Dirichlet boundary condition sets $F(v)=1- v^{D-2}$. 

The physics becomes richer in $D=4-\epsilon$ spacetime dimensions, where the bulk system without boundary flows to its Wilson-Fisher fixed point. Because the bulk system is interacting at low energies, the method of images can no longer be used. Instead, perturbative expansions in $\epsilon$ or $1/n$ may be used to compute $F(v)$. In the following, we take the route of the large-$n$ expansion,  synthesizing the analysis in  Refs.~\cite{Bray:1977tk,Bray:1977fvl,McAvity:1995zd}. Interested readers may look to Refs.~\cite{Bray:1977fvl,Diehl:1981jgg,McAvity:1995zd} for computations using the $\epsilon$-expansion. 

We first make $\varphi_I$ Gaussian by introducing a Hubbard-Stratonovich field, $\lambda$, 
\begin{equation}
    S_{\mathrm{bulk}}=\int d^3r\;\left[ \frac{1}{2}(\partial_{\mu}\varphi_I)^2+\frac{n}{2u}\lambda^2+i\lambda(\varphi_I)^2\right]\,.
\end{equation}
The classical field configurations are the $n\rightarrow\infty$ solutions to the saddle point equations,
\begin{equation}
    (\partial^2-i\lambda_c)\varphi_{Ic}=0,\quad  i\lambda_c=\frac{u}{n}(\varphi_{I })^2\,.
\end{equation}
where the subscript $c$ denotes the classical value (the one-point function). Note that the normal distance $y$ to the boundary is left invariant by the residual conformal symmetry group, so one-point functions of scalar fields are in general not uniform in $y$. Indeed, they generically exhibit universal scaling behavior in $y$,  
\begin{equation}
    \varphi_{Ic}=\frac{a_{\varphi I}}{y^{\Delta\varphi}},\quad  i\lambda_c= \frac{\mu^2-1/4}{y^{2}},
    \label{eq: app lambda ansatz}
\end{equation}
where $a_{\varphi I}$ and $\mu$ are universal constants. The shift of $\mu^2$ by $1/4$ is a common convention in the boundary criticality literature, and the scaling exponent of $\lambda_c$ is fixed to ${\Delta_\lambda=2}$ in the $n\rightarrow\infty$ limit. 

The value of $\mu$ depends on the choice of boundary conditions. To see this, one can  compute the propagator of $\varphi_{I}$ as a general function of $\mu$ by solving  
\begin{equation}
    \left(\partial^2-\frac{\mu^2-1/4}{y^{2}}\right)G_{\varphi}(\scriptr,\ell)=\delta(r-r')\,.
    \label{eq: app G varphi eq}
\end{equation}
Taken together with the saddle point equation governing $\lambda_c$, one finds that the only self-consistent solutions correspond to special values of $\mu^2$, each of which leads to a propagator $G_\varphi$ respecting a different choice of boundary conditions. 
We refer readers to Refs.~\cite{Bray:1977fvl,Bray:1977tk} for detailed discussion of how $\mu^2$ is computed in general dimensions. Here we simply quote the conclusion. In the $n\rightarrow\infty$ limit, the ordinary fixed point has $\mu^2=0$; the special fixed point has $\mu^2=1/4$, meaning $\lambda_c=0$; and $\mu^2=1$ at the extraordinary fixed point. The inclusion of $1/n$ corrections will lead to an RG flow for $\mu^2$, with the fixed point values $\mu^2=0,1$ remaining stable, but $\mu^2=1/4$ becoming unstable in $D=3$~\cite{Krishnan:2023cff,Diehl:2020rfx}. For the remainder of this Appendix, we shall focus on the stable fixed points, the ordinary and the extraordinary.

Equipped with the understanding that different values of $\mu^2$ correspond to the different boundary universality classes, we may solve Eq.~(\ref{eq: app G varphi eq}) for $F(v)$ in the $n\rightarrow\infty$ limit for $D=3$ dimensions. Note that for the extraordinary case, we will assume the saddle point value of $\varphi$ is aligned as $\varphi_{c,1}\neq0$, with its other components vanishing. We therefore ignore the fluctuations of $\varphi_{1}$ about its saddle point value (since they are gapped) and only consider its fluctuations in the gapless, broken symmetry directions.

In the $n\rightarrow\infty$ limit, the order parameter scaling dimension takes its mean field value, $\Delta_\varphi=1/2$. Using this value, one may treat Eq.~\eqref{eq: app prop ansatz} as an ansatz and plug it into Eq.~(\ref{eq: app G varphi eq}) to obtain an ordinary differential equation governing $F(v)$~\cite{Herzog:2020lel},
\begin{equation}
\label{eq: app ODE 0}
    (1-v^2)^2 \frac{d^2F(v)}{dv^2}-(4\mu^2-1)F(v)=0.
\end{equation}
This second order differential equation needs two boundary conditions. The first is the bulk consistency requirement, $F(0)=1$. The second is $F(1)=0$, which requires  $G_{\varphi}$ to vanish on the boundary. 

The requirement, $F(1)=0$, can be understood from the perspective of the boundary OPE. The approach to the boundary is a singular limit, in the sense that bulk one-point functions depend algebraically on the normal coordinate $\sim y^{\Delta}$. Hence, in the boundary limit, bulk operators should be represented by a series of boundary-localized operators (see Refs.~\cite{Gliozzi:2015qsa,billo2016defects,Herzog:2017xha} for an extended discussion). For example, the components of the order parameter field may be expanded as
\begin{align}
    \lim_{y\to0}\varphi\sim \sum_{\widetilde{O}}\,c_{\varphi \widetilde{O}}\,y^{\Delta_{\widetilde{O}}-\Delta_{\varphi}}\widetilde{O},
\end{align}
where the sum is over scalars, $\widetilde{O}$, of the reduced conformal symmetry in the presence of the boundary. Note that we have dropped flavor indices for brevity.  Operators commonly appearing in the boundary OPE are the identity operator (e.g. when the order parameter has an expectation value, as in the case of the extraordinary boundary condition); the derivative operator, $\partial_y\varphi(y=0)$; the displacement operator, $D=T_{yy}(y=0)$, where $T_{\mu\nu}$ is the stress tensor; and the tilt operator $t=J_{y}(y=0)$, which played a starring role in the main text. The particular set of operators, $\widetilde{O}$, furnishing the boundary OPE is determined by boundary conditions of the bulk field.

The boundary OPE leads to the boundary conformal block that characterizes the asymptotic behavior of correlation functions. Looking to the example of the two-point function, 
\begin{align}
    G_{\varphi}(\rho-\rho';y,y'\rightarrow 0)&=\lim_{y,y'\to0}\Big(\langle \varphi(r)\varphi(r')\rangle-\langle\varphi(r)\rangle\langle\varphi(r')\rangle\Big)\\
    &\sim \sum_{\widetilde{O} }|c_{\varphi \widetilde{O}}|^2 (yy')^{\Delta_{\widetilde{O}}-\Delta_{\varphi}}\langle \widetilde{O}(\rho)\widetilde{O}(\rho')\rangle\,,
\end{align}
where we have used the fact that boundary operators with different scaling dimensions are orthogonal (we assume no two scalar boundary operators are degenerate). Since we are computing a connected correlation function, the identity contribution is canceled off on the right-hand-side, leaving only operators with scaling dimension $\Delta_{\widetilde{O}}\geq\Delta_\varphi$. Hence, the right-hand-side contains only regular contributions in $y$ and thus vanishes when $y=y'=0$ unless $\Delta_{\widetilde{O}}=\Delta_{\varphi}$, which is not the case for ordinary and extraordinary boundaries. In terms of  $F(v)$, this conclusion is simply the advertised condition, $F(1)=0$.

The differential equation in Eq.~\eqref{eq: app ODE 0} can be solved with the boundary conditions, $F(0)=1$ and $F(1)=0$, to give
\begin{align}
    F(v)=(1-v)^{1/2+\mu}(1+v)^{1/2-\mu}\,.
\end{align}
Inserting this result into the definition of $G_{\varphi}(\scriptr,\ell)$ in Eq.~\eqref{eq: app prop ansatz}, one finds that for ordinary and extraordinary boundaries,
\begin{align}
    &\text{Ordinary }(\mu=0):\quad G_{\varphi}(\scriptr,\ell)=\frac{\sqrt{1-v^2}}{4\pi |\scriptr|}=\frac{1}{4\pi}\sqrt{\frac{1}{|\scriptr|^2}-\frac{1}{|\ell|^2}}\,,\nonumber\\
    &\text{Extraordinary }(\mu=1):\quad  G_{\varphi}(\scriptr,\ell)= \frac{(1-v)^{3/2}}{4\pi |\scriptr|}\,.
\end{align}
In the extraordinary case, we evaluate the Green's function only in the broken symmetry directions. The form of $F(v)$ for different boundary conditions can in turn be understood through boundary conformal blocks. Taking the limit $y, y'\to 0$ and $v\to 1$, 
\begin{align}
     &\text{Ordinary: }\quad \lim_{y,y'\to 0} G_{\varphi}(\scriptr,\ell)\to \frac{\sqrt{4yy'}}{\rho^2},\nonumber\\
    &\text{Extraordinary: }\quad \lim_{ y, y'\to 0}G_{\varphi}(\scriptr,\ell)\to  \frac{(4 y y')^{3/2}}{4\rho^4}.
    \label{eq: correlation on boundary o(n)}
\end{align}
Thus, in the $n\rightarrow\infty$ limit the lowest dimension operator in $\varphi$'s boundary OPE for the ordinary case has scaling dimension 1, and it is identified with the derivative operator  $\partial_y\varphi$. This can be seen  transparently by noting that with the Dirichlet boundary condition, one may simply Taylor expand the order parameter as $\varphi(y)\sim y\,\partial_y\varphi(y=0)+\dots$.

For the extraordinary case, the leading operator in the boundary OPE has dimension 2, and it is identified with the tilt operator, $t=J_y(y=0)$. Reintroducing indices, one finds that the tilt appears in the boundary OPE channel in the broken symmetry directions (i.e. the putative Goldstones) but of course does not appear in the gapped, unbroken symmetry direction. 
Further analysis~\cite{Bray:1977fvl,Gliozzi:2015qsa} can be used to show that the leading operator appearing in the boundary OPE for the unbroken symmetry direction (beyond the identity) is the displacement operator $D=T_{yy}(y=0)$, which has dimension $\Delta_D=3$ and is thus irrelevant on the boundary.

With the propagator of $\varphi$, one is able to work out the correlation functions of all other operators, including $\lambda$ and the stress tensor $T_{\mu\nu}$, and we shall not review them here but guide readers to Refs.~\cite{McAvity:1995zd,Ohno:1983lma,10.1143/PTP.72.736} for further detailed discussion.

\section{Solving the boundary Schwinger-Dyson equations}
\label{app: extra-ordinary-log ah}
In this Appendix, we provide details of our large-$N$ solution to the Schwinger-Dyson equations~\eqref{eq: original sd equation 1} -- \eqref{eq: original sd equation 2}. 

\subsection{Large-$N$ effective action and fluctuation expansion}

We start by explicitly deriving the real-space Schwinger-Dyson equations for the gauge field, Eqs.~\eqref{eq: original sd equation 1} -- \eqref{eq: original sd equation 2}.

The action for the NCCP$^{N-1}$ model with a boundary at $y=0$ is
\begin{align}
   S&=S_{\mathrm{bulk}}+S_{\mathrm{boundary}}\,,\\
   S_{\mathrm{bulk}}&=\int_{y>0} d^3r\left( \frac{N}{4 e^2}f_{\mu\nu}^2+(D_{\mu}\phi_I)^{\dagger}D^{\mu}\phi_I+\frac{N}{2u}\lambda^2+i\lambda\, \phi_I^{\dagger}\phi_I\right),\; I=1,...,N,
\end{align}
with the Hubbard-Stratonovich field $\lambda$ introduced to model fluctuations of $|\phi_I|^2$. The action $S_{\mathrm{boundary}}$ includes the boundary-localized degrees of freedom -- which can be described in terms of the compact scalar field $\sigma$ introduced in Section~\ref{sec: extra-ordinary NCCP$^{N-1}$ model} -- and any bulk-boundary couplings. 
Note that as discussed in the main text, the boundary degrees of freedom couple to the bulk through the gauge field, $a_\mu$, and that direct couplings to the bulk matter are irrelevant in the RG sense.

Because the action is quadratic in the matter fields, $\phi_I$, we may integrate them out. In the IR limit, $e^2,u\rightarrow\infty$, the resulting effective bulk action is 
\begin{align}
     Z&=\int\mathcal{D}\lambda\mathcal{D}a_\mu\mathcal{D}\sigma \exp\left(-S_{\mathrm{eff}}-S_{\mathrm{boundary}}\right)\,,\\
     S_{\mathrm{eff}}&= \Tr\log\left(-\partial_{\mu}\partial^{\mu}+ia_{\mu}\overrightarrow\partial^{\mu}-ia^{\mu}\overleftarrow \partial_{\mu} +a_{\mu}a^{\mu}+i\lambda\right)\nonumber\\
     &= \Tr\log\left(G_{\phi}^{-1}+ia_{\mu}\overrightarrow\partial^{\mu}-i a^{\mu}\overleftarrow \partial_{\mu} + a_{\mu} a^{\mu}+i\delta \lambda\right),
     \label{eq: Tr log}
\end{align}
where we have introduced the arrow notation on differential operators to denote whether they act to the left or right in the functional trace. In the second line we have decomposed $\lambda=\lambda_{c}+\delta \lambda$ into its classical value and fluctuations, and we have defined $G_{\phi}^{-1}=-\partial^2+\lambda_c(y)$. In the $N\rightarrow\infty$ limit, $\lambda_c(y)$ takes its corresponding large-$N$ value for the O$(2N)$ model, Eq.~\eqref{eq: app lambda ansatz}, with $\mu^2=0$ for the ordinary (Dirichlet) boundary condition. We assume the saddle point value of $a_\mu$ vanishes.

The effective action can be expanded in powers of the fluctuations, $a_\mu$ and $\delta\lambda$, 
\begin{align}
S_{\mathrm{eff}}&=N\,\text{Tr}\log \left(G_{\phi}^{-1}\right) +N\,\text{Tr}\sum_{n=1}^{\infty}\frac{(-1)^{n+1}}{n}\left[G_{\phi}\left(ia_{\mu}\overrightarrow\partial^{\mu}-ia^{\mu}\overleftarrow \partial_{\mu}+ a_{\mu}a^{\mu}+2\delta \lambda\right)\right]^{n}.
\label{eq: app saddle expansion}
\end{align}
So far, the presence of a boundary has seemingly not affected the standard derivation of the large-$N$ effective action. However, the functional trace does contain information about the existence of (ordinary) boundary conditions on $\phi_I$, as did the original path integral.  We therefore should proceed by carrying out the expansion in Eq.~\eqref{eq: app saddle expansion} entirely in real space.

To quadratic order in the fluctuations, which go like $a\sim 1/\sqrt{N}$ and $\delta\lambda\sim 1/\sqrt{N}$, the terms coupling the gauge field to $\delta\lambda$ vanish, so we may approximate ${S_{\mathrm{eff}}\approx S_{\mathrm{gauge}}[a]+S_{\lambda}[\delta\lambda]}$. The saddle point solution for $\lambda$ was discussed above in Appendix~\ref{app: review of o(n) boundary criticality} and is unaffected by gauge fluctuations to leading order in the large-$N$ limit, so we focus here on the gauge field. The non-vanishing terms in the gauge field effective action are 
\begin{align}
   S_{\mathrm{gauge}}&= N\int_r \langle r|-G_{\phi}a_{\mu}a^{\mu}|r\rangle-\frac{N}{2}\int_{r,r'} \langle r|G_{\phi}(ia_{\mu}\overrightarrow\partial^{\mu}-i a^{\mu}\overleftarrow \partial_{\mu})|r'\rangle\langle r'|G_{\phi}(i  a_{\mu}\overrightarrow\partial^{\mu}-i  a^{\mu}\overleftarrow \partial_{\mu})|r\rangle+\mathcal{O}(a^4)\nonumber\\
    &=-\frac{1}{2}\int_{r,r'} a_{\mu}(r)\,\Pi_{\mu\nu}(r,r')\, a_{\nu}(r')\,.
    \label{eq: B6}
\end{align}
Here we have abbreviated $\int_{y>0} d^3r$ to $\int_r$ and defined the one-loop polarization tensor, 
\begin{align}
    \Pi_{\mu\nu}(r,r')=2N\Big[&G_{\phi}(r,r')\,\delta(r-r')\,\delta_{\mu\nu} \nonumber\\
     & \qquad + G_{\phi}(r,r')\partial_{\mu}\partial_{\nu}'G_{\phi}(r,r')-\partial_{\mu}G_{\phi}(r,r')\partial_{\nu}'G_{\phi}(r,r')\Big]\,. 
     \label{eq: app full pi}
\end{align}
Further contributions to the polarization tensor are suppressed by powers of $1/N$. 

The polarization tensor in Eq.~\eqref{eq: B6} must be consistent with gauge invariance, meaning that it must respect the real space Ward identity,
\begin{align}
\label{eq: app Ward pi}
\partial_\mu \Pi_{\mu\nu}(r,r')=\partial'_\nu\Pi_{\mu\nu}(r,r')=0\,,
\end{align}
which follows from immediately from the fact that $G_\phi$ satisfies Eq.~\eqref{eq: app G varphi eq} in the large-$N$ limit. 

The delta function contact term in Eq.~\eqref{eq: B6} -- the ``diamagnetic'' contribution -- is essential for guaranteeing that the Ward identity is respected locally as $r\rightarrow r'$. However, the true importance of this term depends on the short-distance regularization, and its contribution to the real space Schwinger-Dyson equations can be chosen to vanish. This corresponds to computing
\begin{align}
G_\phi(r,r)=\lim_{y\rightarrow y'}\int\frac{d^2k}{(2\pi)^2}\,\widetilde{G}_\phi(k;y,y')\,,
\end{align}
where $\widetilde{G}_\phi(k;y,y')$ is the Fourier transform of the matter propagator in the plane parallel to the boundary. By analytic continuation of the boundary co-dimension, $\tilde{d}=2+\epsilon$, this integral may be evaluated and shown to vanish~\cite{Bray:1977fvl}. This is analogous to the case without a boundary, where the diamagnetic contribution leads to UV divergences that vanish under dimensional regularization.

\subsection{Schwinger-Dyson equations}

To construct the Schwinger-Dyson equations for the gauge field, we choose a gauge in which fluctuations are transverse to the plane of the boundary,
\begin{align}
\partial_ma_m=0\,,
\label{eq: app transverse gauge}
\end{align}
where we recall the use of Latin indices for boundary coordinates. 

We wish to implement this gauge by introducing an extra gauge fixing term to the action, 
\begin{align}
    S_{\mathrm{gauge}}=-\frac{1}{2}\int_{r,r'}\, a_{\mu}(r)\Big[\Pi_{\mu\nu}(r,r')+\mathcal{G}_{\mu\nu}(r,r')\Big]a_{\nu}(r')+...\,,
    \label{eq: app gauge action G}
\end{align}
where the gauge fixing term is defined as
\begin{align}
    \int_{r,r'}a_\mu(r)\,\mathcal{G}_{\mu\nu}(r,r')\,a_\nu(r')&=-\frac{N}{2(\zeta-1)}\int_{r,r'}[\partial_m a_m(r)]\,\mathcal{K}(r,r')\,[\partial_n a_n(r')]\,,
\end{align}
where $\zeta$ is a dimensionless parameter. The gauge \eqref{eq: app transverse gauge} corresponds to the constraint imposed by the limit $\zeta\rightarrow 1$. We assume that a non-local kernel, ${\mathcal{K}(r,r')=\mathcal{K}(\rho-\rho',y,y')}$, exists such that the IR limit $e^2\rightarrow\infty$ used to drop the Maxwell term may be taken prior to the limit $\zeta\rightarrow 1$, analogous to the boundary-free case~\cite{Chester:2016ref}. In principle, one may wish to construct the detailed form of $\mathcal{K}$ to obtain a full family of gauges parameterized by $\zeta$. However, for our purposes we will need only to make use of the fact that in the $\zeta\rightarrow 1$ limit the photon propagator, $D_{\mu\nu}(r,r')=\langle a_\mu(r)a_\nu(r')\rangle$, must be transverse,
\begin{align}
\partial_m D_{m\mu}=\partial_m D_{\mu m}=0\,.
\end{align}
and have vanishing off-diagonal components,
\begin{align}
D_{ym}=D_{my}=0\,,
\label{eq: app vanishing Dym}
\end{align}
per the argument in Section~\ref{subsec: saddle point eqs} of the main text. With these conditions applied self-consistently, the detailed form of $\mathcal{K}$ will drop out of the final Schwinger-Dyson equations, as we now demonstrate.

From the quadratic action in Eq.~\eqref{eq: app gauge action G}, we observe that the gauge propagator in the $N\rightarrow\infty$ limit is the solution to the integral equation,
\begin{align}
    \int_w\,\Big[\Pi_{\mu\rho}(r,w)+\mathcal{G}_{\mu\rho}(r,w)\Big]D_{\rho\nu}(w,r')&=-\delta_{\mu\nu}\,\delta(r-r')\,,
    \label{eq: sd equation 1}
\end{align}
with $\Pi_{\mu\nu}$ given in Eq.~\eqref{eq: app full pi}. Using Eq.~\eqref{eq: app vanishing Dym}, we see this equation can be split into independent block components,
\begin{align}
    \label{eq: app sd equation 1}
    \int_w\,\Big[\Pi_{ml}(r,w)+\mathcal{G}_{ml}(r,w)\Big]D_{ln}(w,r')&=-\delta_{mn}\,\delta(r-r')\,,\\
    \int_w\,\Pi_{yy}(r,w)D_{yy}(w,r')&=-\delta(r-r')\,,
    \label{eq: sd equation 2}
\end{align}
where we have also used the fact that $\mathcal{G}_{\mu y}=\mathcal{G}_{y\mu}=0$ by definition. Taking the divergence of both sides of Eq.~\eqref{eq: app sd equation 1} with respect $r_m$ and using the Ward identity, Eq.~\eqref{eq: app Ward pi},
\begin{align}
    -\partial_{n}\delta(r-r')=&\int_w\,\Big[\partial_m\Pi_{ml}(r,w)+\partial_{m}\mathcal{G}_{ml}(r,w)\Big]D_{ln}(w,r')\nonumber\\
    =&\int_w\,\Big[-\partial_y\Pi_{yl}(r,w)+\partial_{m}\mathcal{G}_{ml}(r,w)\Big]D_{ln}(w,r')\nonumber\\
    =&\int_w\,\partial_{m}\mathcal{G}_{ml}(r,w)D_{ln}(w,r')\,.
    \label{eq: sd equation 3}
\end{align}
The first term in the second line vanishes by the $\mu=y$ components of Eq.~\eqref{eq: sd equation 1}. We therefore infer that for $\zeta\rightarrow 1$, $\mathcal{G}_{mn}$ and $D_{mn}$ satisfy the integral equation,
\begin{align}
    -\frac{\partial_{m}\partial_{n}}{\partial_l\partial_l}\,\delta(r-r')=\int_w\,\mathcal{G}_{ml}(r,w)D_{ln}(w,r')\,.
\end{align} 
Subtracting this equation from Eq.~(\ref{eq: sd equation 1}), we obtain our final set of Schwinger-Dyson equations for the photon propagator,
\begin{align}
\label{eq: sd equation 4}
\int d^3w\:\Pi_{ml}(r,w)D_{ln}(w,r')&=-\left(\delta_{mn}-\frac{\partial_{m}\partial_{n}}{\partial_l\partial_l}\right)\delta(r-r')\,,\\
\label{eq: sd equation 5}
    \int d^3w\,\Pi_{yy}(r,w)D_{yy}(w,r')&=-\delta(r-r')\,.
\end{align}
The gauge fixing term $\mathcal{G}_{ml}$ is thus removed from the original Schwinger-Dyson equation, with its effect reflected by the transverse tensor on the RHS of Eq.~(\ref{eq: sd equation 4}). One can check that Eq.~(\ref{eq: sd equation 4}) indeed vanishes upon taking divergence with respect $r_n'$ of both sides. 

Our goal is now to solve these equations given the saddle point solution for the polarization tensor, Eq.~\eqref{eq: app full pi}.  

\subsection{``Diagonalizing'' the Schwinger-Dyson equations}

We start by focusing on Eq.~\eqref{eq: sd equation 4}. After gauge fixing, $\partial_mD_{m\mu}=0$, not all components of $\Pi_{mn}$ contribute to the integral Schwinger-Dyson equations~\eqref{eq: sd equation 4} -- \eqref{eq: sd equation 5}. Indeed, a  total derivative term may always be added to $\Pi_{mn}$ without changing the solution for $D_{mn}$, 
\begin{align}
    \int_w\left[ \frac{\partial}{\partial w_m}B_n(r,w)\right] D_{mj}(w,r')&= 0,
\end{align}
for a general vector-valued function, $B_n(r,w)$. Therefore, $\Pi_{mn}$ may be ``diagonalized'' using a suitable choice of $B_n$. 

To motivate our choice for $B_n$, we begin by writing down the components of the polarization tensor explicitly, dropping the delta function contact term (which has vanishing contribution to the Schwinger-Dyson equations),
\begin{align}
    \label{eq: app Pimn explicit}
    \Pi_{mn}&=\frac{2N}{(4\pi)^2}\left(\frac{1}{\scriptr^2}-\frac{1}{\ell^2}\right) \Big( \frac{\mathcal{T}_{mn}(\scriptr)}{\scriptr^2}+\frac{\mathcal{T}^R_{mn}(\ell)}{\ell^2}\Big),\,\\
    \Pi_{yy}&=\frac{2N}{(4\pi)^2}\left(\frac{1}{\scriptr^2}-\frac{1}{\ell^2}\right) \Big( \frac{\mathcal{T}_{yy}(\scriptr)}{\scriptr^2}-\frac{\mathcal{T}_{yy}^R(\ell)}{\ell^2}\Big),\\
    \Pi_{m y}&=\frac{2N}{(4\pi)^2}\left(\frac{1}{\scriptr^2}-\frac{1}{\ell^2}\right) \Big( \frac{\mathcal{T}_{my}(\scriptr)}{\scriptr^2}-\frac{\mathcal{T}_{my}^R(\ell)}{\ell^2}\Big)\,,\\
    \Pi_{y m}&=\frac{2N}{(4\pi)^2}\left(\frac{1}{\scriptr^2}-\frac{1}{\ell^2}\right) \Big( \frac{\mathcal{T}_{ym}(\scriptr)}{\scriptr^2}+\frac{\mathcal{T}_{y m}^R(\ell)}{\ell^2}\Big),
    \label{eq: current-current correlation function}
\end{align}
where
\begin{align}
    \mathcal{T}_{\mu\nu}(\scriptr)=\delta_{\mu\nu}-2\frac{\scriptr_{\mu}\scriptr_{\nu}}{\scriptr^2},\quad
    \mathcal{T}^R_{\mu\nu}(\ell)=\delta_{\mu\nu}-2\frac{\ell_{\mu}\ell_{\nu}}{\ell^2}.
\end{align}
and absolute value signs on $\scriptr^2$ and $\ell^2$ are left implicit from here. Inspecting Eq.~\eqref{eq: app Pimn explicit}, one may expect that a function of the form,
\begin{align}
\partial_m B_n=\frac{N}{8\pi^2}\,\partial_m\partial_n\left(\frac{b(v)}{\scriptr^2}\right)\,,
\end{align}
may be subtracted from $\Pi_{mn}$ to render it diagonal. Here $b(v)$ is a scalar function of the crossing ratio, $v=|\scriptr|/|\ell|$, which should satisfy
\begin{equation}
  \partial_m\partial_n\left(\frac{b(v)}{\scriptr^2}\right)=-2\left(\frac{1}{\scriptr^2}-\frac{1}{\ell^2}\right)\Big(\frac{\scriptr_m\scriptr_n}{\scriptr^4}+\frac{\ell_m\ell_n}{\ell^4}\Big)+f(\scriptr,\ell)\,\delta_{mn}\,,
  \label{eq: equation about b}
\end{equation}
where $f(\scriptr,\ell)$ is a function determined by $b$ and its derivatives.  Indeed, subtracting $\partial_m B_n$ from $\Pi_{mn}$, we obtain a diagonal tensor at the expense of introducing $f(\scriptr,\ell)$,
\begin{equation}
    \Pi_{mn}-\partial_mB_n  =\frac{N}{8\pi^2}\left(\frac{1}{\scriptr^4}-\frac{1}{\ell^4}-f(\scriptr,\ell)\right)\,\delta_{mn}\,.
\end{equation}
In the following, we solve for $b(v)$. To obtain a unique solution, it will suffice to require that $b(v)$ and its derivatives are finite in the boundary limit, $v\rightarrow 1$.   

To establish the relationship between $b$ and $f$, we expand the LHS of Eq.~\eqref{eq: equation about b},
\begin{align}
    \partial_m\partial_n\left(\frac{b}{\scriptr^2}\right)=&\left[\left(\frac{1}{\scriptr\ell}-\frac{\scriptr}{\ell^3}\right)^2\frac{d^2b}{dv^2}
    + \left(-\frac{5}{\scriptr^3\ell}+\frac{2}{\scriptr\ell^3}+3\frac{\scriptr}{\ell^5}\right)\frac{db}{dv}+8\frac{b}{\scriptr^4}\right]\frac{\scriptr_m\scriptr_n}{\scriptr^2}
    \nonumber\\
    &+\left[\left(\frac{1}{\scriptr^3\ell}-\frac{1}{\scriptr\ell^3}\right)\frac{db}{dv}-\frac{2}{\scriptr^4}b\right]\delta_{mn}\,.
    \label{eq: partial_mpartial_nb/r^2 expanded}
\end{align}
where we have used the fact that $\scriptr_m=\ell_m$ to simplify the tensor structure on the first line. By inspection, we see that the second line defines $f(\scriptr,\ell)$. Rewriting it in terms of the crossing ratio,
\begin{align}
    f(\scriptr,\ell) =\frac{1}{\scriptr^4}\Big((v-v^3)\frac{db}{dv}-2b\Big)\,.
    \label{eq: definition of f1}
\end{align}
To make further progress, it will be useful to break up the differential equation~\eqref{eq: equation about b} into several parts by decomposing
\begin{align}
b(v)=b_1(v)+b_2(v)+b_3(v)\,,
\end{align}
where
\begin{align}
\label{eq: b1}
\partial_m\partial_n\left(\frac{b_1(v)}{\scriptr^2}\right)&=-2\frac{\scriptr_m\scriptr_n}{\scriptr^6}+2\frac{\ell_m\ell_n}{\ell^6} +f_1(\scriptr,\ell)\delta_{mn}\,,\\
    \label{eq: b2}
\partial_m\partial_n\left(\frac{b_2(v)}{\scriptr^2}\right)&=2\frac{\scriptr_m\scriptr_n}{\scriptr^4\ell^2}+f_2(\scriptr,\ell)\delta_{mn}\,,\\
\label{eq: b3}
\partial_m\partial_n\left(\frac{b_3(v)}{\scriptr^2}\right)
    &= -2\frac{\ell_m\ell_n}{\scriptr^2\ell^4} +f_3(\scriptr,\ell)\delta_{mn}\,,
\end{align}
and $f_{1},f_2,f_3$ respectively satisfy Eq.~\eqref{eq: definition of f1} for $b_1,b_2,b_3$. 

By guessing 
\begin{align}
f_1(\scriptr,\ell)&=\frac{1}{2}\left(\frac{1}{\scriptr^4}-\frac{1}{\ell^4}\right)\,,
\end{align}
we may obtain a simple solution for $b_1$, 
\begin{align}
    b_1(v)&=-\frac{1}{4}\,(1-v^2)\,,
\end{align}
which is indeed finite in the limits $v\rightarrow0$ and $v\rightarrow 1$. Using the expansion in Eq.~(\ref{eq: partial_mpartial_nb/r^2 expanded}) and matching terms proportional to $\scriptr_m\scriptr_n$, it is straightforward to recast Eq.~\eqref{eq: b2} as a single ordinary differential equation for $b_2(v)$, 
\begin{align}
    \left(1-v^2\right)^2\frac{d^2b_2}{dv^2} + \left(-\frac{5}{v} + 2v + 3v^3\right)\frac{db_2}{dv} + 8\frac{b_2}{v^2} = 2\,,
\end{align}
which has the general solution,
\begin{equation}
    b_2(v) = - \frac{v^2(1+2C_1(v^2-1)-2C_2 v^2 + \log v^2)}{2(v^2-1)^2}\,,
    \label{eq: b2v in general}
\end{equation}
where $C_1,C_2$ are constants determined by boundary conditions. Requiring that $b_2(v)$ be finite as $v\rightarrow 1$ uniquely fixes $b_2(v\rightarrow1)=1/4$ and 
\begin{equation}
    C_1=0\,,\quad C_2=\frac12\,,
\end{equation}
leading to a particular solution,
\begin{align}
    b_2(v)= - \frac{v^2(1-v^2)+v^2\log v^2}{2(v^2-1)^2}\,,\quad f_2(\scriptr,\ell)=\frac{v^2}{\scriptr^4}\left(\frac{v^2-1-v^2\log v^2}{(v^2-1)^2}\right).
\end{align}
The same strategy can be applied to obtain a differential equation for $b_3$,
\begin{align}
    (1-v^2)^2\frac{d^2b_3}{dv^2} + \left(-\frac{5}{v} + 2v + 3v^3\right)\frac{db_3}{dv} + 8\frac{b_3}{v^2} = -2v^2,
\end{align}
where finiteness of the solution as $v\rightarrow 1$ similarly mandates $b_3(v\rightarrow 1)=-1/4$. The unique solution is
\begin{equation}
   b_3(v)=- \frac{v^2(1-v^2)+v^4\log v^2}{2(v^2-1)^2},\quad f_3(\scriptr,\ell)=\frac{v^4}{\scriptr^4}\frac{(v^2-1-\log v^2)}{(v^2-1)^2}\,.
\end{equation}

Putting these results together, we obtain 
\begin{align}
    F(\scriptr,\ell)\delta_{mn}\equiv\Pi_{mn}
    -\frac{N}{8\pi^2}\partial_m\partial_n\frac{\sum_{i=1}^3b_i}{\scriptr^2}
    =\frac{N}{(4\pi)^2}\frac{\delta_{mn}}{\scriptr^4}
    \left(\frac{1-v^8+4v^4\log v^2}{(v^2-1)^2}\right),
\end{align}
which allows us to take the trace of Eq.~(\ref{eq: sd equation 4}), leading to Eq.~(\ref{eq: FD equation}) of the main text,
\begin{equation}
  \int d^3w\,F (r,w)\mathfrak{D}(w,r')= -\delta(r-r'),
   \label{eq: Scwinger-Dyson finally 1}
\end{equation}
where
\begin{align}
    \mathfrak{D}(w,r')=D_{\tau\tau}(w,r')+D_{xx}(w,r')\,.
\end{align}

\subsection{Photon propagator}
The reduced Schwinger-Dyson Eq.~(\ref{eq: Scwinger-Dyson finally 1}) can be solved using methods developed in Ref.~\cite{McAvity:1995zd}. Although the physical meaning of the crossing ratio $v=\scriptr/\ell$ is transparent, it will be  convenient to introduce an additional dimensionless crossing ratio,
\begin{align}
   \xi = \frac{\scriptr^2}{\ell^2-\scriptr^2}
   = \frac{\scriptr^2}{4r_yw_y}
   = \frac{v^2}{1 - v^2}\,.
\end{align}
Rewriting $F$ as a function solely of $\xi$, 
\begin{align}
   F = \frac{N}{(4r_yw_y)^2}h(\xi), \quad  h(\xi) 
   =\frac{1}{(4\pi)^2}\Big( \frac{1}{\xi^2(1+\xi)^2}
   + \frac{4}{\xi}
   - \frac{2}{(1+\xi)^2}
   + 4\log\frac{\xi}{1+\xi}\Big).
\end{align}

The trace of the photon propagator, $\mathfrak{D}$, may also be defined in terms of a dimensionless function of the crossing ratio, $\xi'$, where $r$ is substituted with $r'$,
\begin{align}
    \mathfrak{D}(w,r')=\frac{1}{4w_yr'_y}\lambda(\xi'),\quad  \xi' 
   = \frac{\scriptr'^2}{4w_yr_y'}
   = \frac{(v')^2}{1 - (v')^2}.
\end{align}
This ansatz is motivated by the action of the residual conformal symmetry group, under which $J^{y}_{\mathrm{top}}$ (the electric field of $a_\mu$) must have scaling dimension two. That $\mathfrak{D}$ behaves like a scalar under the residual conformal symmetry follows from our choice of gauge, and we will indeed see that this ansatz leads to a self-consistent solution. Note that the function, $\lambda$, here is not to be confused with the Hubbard-Stratonovich field used elsewhere in the text.

Integrating over parallel coordinates of $r$, which we denote as $\rho$, the Schwinger-Dyson equation further reduces to 
\begin{align}
    \int d^3w \int d^2\rho \,F(r,w)\frac{1}{4w_yr'_y}\lambda(\xi')=-\delta(r_y-r_y')\,.
\end{align}
With the expression of $F$, one has
\begin{align}
    \int d^2\rho\, F(r,w)&=\frac{N\pi}{4r_yw_y}\int_0^{\infty} d(\frac{\rho^2}{4r_yw_y})\,h\Big(\frac{\rho^2}{4r_yw_y}+\frac{(r_y-w_y)^2}{4r_yw_y}\Big)\nonumber\\
    &=\frac{N\pi}{4r_yw_y}\frac{2}{(4\pi)^2}\Big(-2-(1+2\chi)\log\frac{\chi}{1+\chi}+\frac{1}{2\chi(1+\chi)}\Big)\nonumber\\
    &=\frac{N\pi}{4r_yw_y}h(\chi),
\end{align}
where 
\begin{align}
   \chi=\frac{(r_y-w_y )^2}{4r_yw_y }.
\end{align}
is a third species of dimensionless ratio. Since $N\pi h(\chi )/4r_y w_y$ is independent of $w_\tau,w_x$, one can formally integrate $\lambda(\xi)$ over $w_\tau,w_x$ and denote the result as 
\begin{align}
    \int d^2w\;  \frac{1}{4w_yr'_y}\lambda(\xi')=\pi \Lambda(\chi')\,,\qquad \chi'=\frac{(r'_y-w_y )^2}{4r'_yw_y }\,,
\end{align} 
where $\Lambda(\chi')$ is a dimensionless function of $\chi'$. The Schwinger-Dyson equation is now recast in a simpler form,
\begin{align}
    \int_0^{\infty}dw_y\, \frac{\pi^2}{4r_yw_y }h(\chi)\Lambda(\chi')=-\frac{1}{N}\delta(r_y-r_y').
\end{align}
Now we define new coordinates
\begin{align}
    &w_y=e^{2\theta},\quad r_y=e^{2\alpha},\quad r_y'=e^{2\alpha'},
\end{align}
such that
\begin{align}
    \chi=\left(\frac{e^{\alpha-\theta}-e^{\theta-\alpha}}{2}\right)^2=\sinh^2(\theta-\alpha)&, \:\chi'=\left(\frac{e^{\theta-\alpha'}-e^{\alpha'-\theta}}{2}\right)^2=\sinh^2(\theta-\alpha'),
\end{align}
and the above integral equation becomes 
\begin{align}
    &\int_{-\infty}^{\infty}d\theta \:h(\sinh^2(\theta-\alpha))
    \lambda(\sinh^2(\theta-\alpha'))= -\frac{1}{N \pi^2}\delta(\alpha-\alpha').
\end{align}
In particular, we will leverage the Fourier transform with respect to $\theta$, 
\begin{equation}
    \hat{h}(k)=\int_{-\infty}^{\infty} d\theta \: e^{i\theta k}h(\sinh^2\theta),
\end{equation}
and similarly for $\hat{\lambda}(k)$. The transformed Schwinger-Dyson equation is 
\begin{align}
    \int \frac{dk}{2\pi}\,\hat{h}(k)\hat{\lambda}(k)\,e^{ik(\alpha-\alpha')}=-\frac{1}{N\pi^2}\delta(\alpha-\alpha')\,.
\end{align}
The solution is simply $\hat{\lambda}(k)=(-N\pi^2\hat{h}(k))^{-1}$. 

In the following, we work out $\hat{h}(k)$, from which we can extract position space solutions via inverse Fourier transform.  
The strategy is to utilize Hypergeometric functions, which have well-defined Fourier transform. Introducing a new function,
\begin{align}
    &g_{a,b}(\chi)=\frac{\Gamma(2a)}{4^{2a-1}\Gamma(b-a)\Gamma(b+a)}\frac{1}{(1+\chi)^{2a}}{}_2F_1(2a,a+b-\frac{1}{2};2a+2b-1;\frac{1}{1+\chi}),
    \label{eq: g a b chi}
\end{align}
where ${}_2F_1$ is the Gauss Hypergeometric function, defined as
\begin{align}
    {}_2F_1(a,b;c;x)=\sum_{n=0}^{\infty}\frac{(a)_n(b)_n}{(c)_n}\frac{x^n}{n!},
\end{align}
where $(a)_n$ is the Pochhammer symbol,
\begin{align}
    \quad (a)_n=\begin{cases}
        1 & n=0\\
        a(a+1)...(a+n-1)&n>0.
    \end{cases}
\end{align}

By the expression of $h(\chi)$,
\begin{align}
    h(\chi)=&\frac{2}{(4\pi)^2}\Big(-2-(1+2\chi)\log\frac{\chi}{1+\chi}+\frac{1}{2\chi(1+\chi)}\Big)\nonumber\\
    =&\frac{2}{(4\pi)^2}\Big(-2\pi g_{1,\frac{1}{2}}(\chi)+\frac{\pi}{2}g_{1,\frac{3}{2}}(\chi)\Big),
\end{align}
where
\begin{align}
g_{1,\frac{1}{2}}(\chi)=-\frac{1}{4\pi}\frac{1}{\chi(1+\chi)},\quad g_{1,\frac{3}{2}}(\chi)=-\frac{2}{\pi}(2+(1+2\chi)\log\frac{\chi}{1+\chi}).
\end{align}
Based on Appendix B of Ref.~\cite{McAvity:1995zd}, the Fourier transform of $g_{a,b}(\chi)$ has a compact expression,
\begin{align}
    \hat{g}_{a,b}(k)=\int_{-\infty}^{\infty} d\theta\; e^{ik\theta} g_{a,b}(\sinh^2\theta)=\frac{\Gamma(a-ik/4)\Gamma(a+ik/4)}{\Gamma(b-ik/4)\Gamma(b+ik/4)}.
\end{align}
Therefore, 
\begin{align}
    \hat{h}(k)=&\frac{2}{(4\pi)^2}(-2\pi \hat{g}_{1,\frac{1}{2}}(k)+\frac{\pi}{2}\hat{g}_{1,\frac{3}{2}}(k))\nonumber\\
    =&\frac{2}{(4\pi)^2}\Big(-2\pi\frac{\Gamma(1-ik/4)\Gamma(1+ik/4)}{\Gamma(1/2-ik/4)\Gamma(1/2+ik/4)}+\frac{\pi}{2}\frac{\Gamma(1-ik/4)\Gamma(1+ik/4)}{\Gamma(3/2-ik/4)\Gamma(3/2+ik/4)}\Big)\nonumber\\
    =&-\frac{k^2}{64\pi}\frac{\Gamma(1-ik/4)\Gamma(1+ik/4)}{\Gamma(3/2-ik/4)\Gamma(3/2+ik/4)} .
\end{align}
The inverse of it is obtained by simply swapping the denominator and the numerator,
\begin{align}
    \hat{\lambda}(k)=(-N\pi^2\hat{h}(k))^{-1}=\frac{64}{\pi  k^2}\frac{\Gamma(3/2-ik/4)\Gamma(3/2+ik/4)}{\Gamma(1-ik/4)\Gamma(1+ik/4)}=\frac{64}{N\pi  k^2}\hat{g}_{\frac{3}{2},1}(k),
\end{align}
whose Fourier transform picks up a pole at $k=0$, 
\begin{align}
    &\int \frac{dk}{2\pi}\, e^{-ik\theta}k^2\hat{\lambda}(k)=-\frac{d^2}{d\theta^2}\lambda(\sinh^2\theta)=\frac{64}{N\pi  }g_{\frac{3}{2},1}(\sinh^2\theta).
\end{align}
By the definition $\chi'=\sinh^2\theta$, one has
\begin{align}
    &\frac{d^2}{d\theta^2}\Lambda(\chi')=4\chi'(1+\chi')\frac{d^2}{d\chi'^2}\Lambda(\chi') +2(1+2\chi')\frac{d}{d\chi'}\Lambda(\chi'),
\end{align}
thus
\begin{align}
    4\chi'(1+\chi')\frac{d^2}{d\chi'^2}\Lambda(\chi') +2(1+2\chi')\frac{d}{d\chi'}\Lambda(\chi')=-\frac{64}{N\pi  }g_{\frac{3}{2},1}(\chi').
    \label{eq:master equation of d}
\end{align}
One can solve $\Lambda(\chi')$ from this differential equation. But a further observation suggests that this equation can be converted directly to that about $\lambda(\xi')$. By definition,
\begin{align}
    &\int d^2w \,\frac{1}{4w_yr'_y} \lambda(\xi')=\pi \Lambda(\chi'),
\end{align}
and $\xi'=\rho'^2/4w_y r'_y+\chi'$, one has
\begin{align}
     \Lambda(\chi')= &\int_0^{\infty} d(\frac{\rho'^2}{4w_y r'_y}) \,\lambda (\frac{\rho'^2}{4w_y r'_y}+\chi')
     \equiv \int_{u'} \lambda(u'+\chi')=\int_{u'}\lambda(\xi'),
\end{align}
here we abbreviated $\int_0^{\infty}du'=\int_{u'}$. Consequently, the following relation can also be derived,
\begin{align}
   \frac{d\,\Lambda(\chi')}{d\chi'}
   =&\int_{u'}\frac{d\,\Lambda(\chi'+u')}{d(\chi'+u')}
   =\int_{u'}\frac{d\,\lambda(\xi')}{d\xi'},
\end{align}
and similarly, 
\begin{align}
  \chi'\,\frac{d\,\Lambda(\chi')}{d\chi'}=&\int_{u'}\xi'\,\frac{d\,\lambda(\xi')}{d\xi'}+\lambda(\xi'),
  \nonumber\\
  \chi'\,\frac{d^2 \Lambda(\chi')}{d{\chi'}^2}
  =& \int_{u'}\xi'\,\frac{d^2 \lambda(\xi')}{d{\xi'}^2}
  +\frac{d\,\lambda(\xi')}{d\xi'},
  \nonumber\\
  \chi'^2\,\frac{d^2 \Lambda(\chi')}{d{\chi'}^2}
  =& \int_{u'}\xi'^2\,\frac{d^2 \lambda(\xi')}{d{\xi'}^2}
  +2\xi'\,\frac{d\,\lambda(\xi')}{d\xi'}.
\end{align}
With these equations, the LHS of~(\ref{eq:master equation of d}) becomes,
\begin{align}
&4\chi'(1+\chi')\frac{d^2}{d\chi'^2}\Lambda(\chi') +2(1+2\chi')\frac{d}{d\chi'}\Lambda(\chi')\nonumber\\
   =& \int_{u'}  4\xi'(1+\xi')\frac{d^2}{d\xi'^2}\lambda(\xi')
   +(6+12\xi')\frac{d}{d\xi'} \lambda(\xi')
   +4\lambda(\xi').
\end{align}

The RHS of~(\ref{eq:master equation of d}) can be transformed similarly, 
\begin{align}
    g_{\frac{3}{2},1}(\chi')= \int_{u'} g_{\frac{3}{2},1}(\xi')
    =\int_{u'} g_{\frac{3}{2},1}(u'+\chi'),
\end{align}
where
\begin{align}
    g_{a,b}(\xi')=\frac{\Gamma(2a+1)}{4^{2a-1}\Gamma(b-a)\Gamma(b+a)}\frac{1}{(1+\xi')^{2a+1}}{}_2F_1(2a+1,a+b-\frac{1}{2},2a+2b-1,\frac{1}{1+\xi'}).
\end{align}
One can check this by integrating $g_{a,b}(\xi')$ over tranverse coordinates, and the result recovers $g_{a,b}(\chi')$ defined in Eq.~(\ref{eq: g a b chi}). Based on its form, one has
\begin{equation}
    g_{\frac{3}{2},1}(\xi')=-\frac{1}{4\pi}\frac{1}{\xi'^2(1+\xi')^2}.
\end{equation}
Therefore,
\begin{align}
    4\xi'(1+\xi')\frac{d^2}{d\xi'^2}\lambda(\xi')+(6+12\xi')\frac{d}{d\xi'}\lambda(\xi') +4\lambda(\xi')=\frac{16}{N\pi^2}\frac{1}{\xi'^2(1+\xi')^2},
\end{align}
of which the general solution is
\begin{align}
    \lambda(\xi')
    =\frac{4}{N\pi^2\sqrt{\xi'^3(1+\xi')}}\Biggl[&
2\sqrt{\frac{\xi'}{1+\xi'}}
-\pi\,\xi'\big(
-\pi
+2\pi^4 C_{1}+4i\log2+4\pi^4C_{2}\log (\sqrt{1+\xi'}+\sqrt{\xi'})\big)\nonumber\\
&+2\xi'\,\mathrm{Li}_{2}(\frac{1}{(\sqrt{\xi'}+\sqrt{1+\xi'})^4})-8\xi'\,\mathrm{Li}_{2}(\frac{1}{(\sqrt{\xi'}+\sqrt{1+\xi'})^2})
\Biggr],
\end{align}
where $C_1,C_2$ are two constants to be determined. Note that the arguments of the polylogrithmic functions $\text{Li}_2$ are always smaller than or equal to 1 given that $\xi'$ is real, so no analytical continuation is required. 

Now let's solve $C_1$ and $C_2$. Note that the propagator should be real, so
\begin{equation}
    C_1=-i\frac{2}{\pi^4}\log2.
\end{equation}
Meanwhile, in the limit to the boundary, $\xi'\to\infty$, 
\begin{align}
    \lim_{\xi'\to\infty}\lambda(\xi')=&4 \frac{-\pi+4C_2\pi^4\log2+2C_2\pi^4\log \xi'}{- N\pi \xi'}+\mathcal{O}(\xi'^{-2}).
\end{align}
Note that the full propagator has an extra $1/4w_yr'_y$ denominator, which makes $d(\xi')$ well defined on the boundary provided the singular term $\log \xi'$ disappears, so we set $C_2=0$. 

Finally, putting everything together, 
\begin{align}
    \lambda(\xi')=
\frac{4}{N\pi^2\sqrt{\xi'+\xi'^2}}\Biggl[
&2\sqrt{\frac{1}{\xi'+\xi'^2}}
+\pi^2+2 \mathrm{Li}_{2}\left(\frac{1}{(\sqrt{\xi'}+\sqrt{1+\xi'})^4}\right)
-8\mathrm{Li}_{2}\left(\frac{1}{(\sqrt{\xi'}+\sqrt{1+\xi'})^2}\right)
\Biggr].
\end{align}
From now on, we can drop the prime superscript since the Schwinger-Dyson equation has been solved, so there is no need to keep tracking $r$ and $r'$ variables.

By the definition, 
\begin{align}
    \mathfrak{D}&=\frac{1}{4w_yr_y}\lambda(\xi)\nonumber\\
&=
\frac{4}{N\pi^2|\scriptr||\ell|}\Biggl[\frac{2(\ell^2-\scriptr^2)}{|\scriptr||\ell|}
+\pi^2+2 \operatorname{\mathrm{Li}_{2}}\left(\frac{(\ell^2-\scriptr^2)^2}{(|\scriptr|+|\ell|)^4}\right)-8 \operatorname{\mathrm{Li}_{2}}\left(\frac{\ell^2-\scriptr^2}{(|\scriptr|+|\ell|)^2}\right)
\Biggr]\nonumber\\
&=\frac{4v}{N\pi^2\scriptr^2}\Biggl[\frac{2(1-v^2)}{v}
+\pi^2+2 \mathrm{Li}_{2}\Big(\frac{(1-v)^2}{(1+v)^2}\Big)-8\mathrm{Li}_{2}\Big(\frac{1-v}{1+v}\Big)
\Biggr]\,,
\label{eq: D11+D22}
\end{align}
which is the result in the main text Eq.~(\ref{eq: trace D result}).

As a sanity check, we can compare our result with the no-boundary result. In the coincident limit $\scriptr\to 0$, the boundary effect should be ignorable and the no-boundary result should be recovered at the leading order. Taking the such coincident limit $r\to r', \xi\to0$, the leading singularity in $\lambda(\xi)$ gives
\begin{align}
   \lim_{\scriptr\to 0}\mathfrak{D}(\scriptr)=\frac{8}{N\pi^2}\frac{1}{\scriptr^2}.
   \label{eq: no boundary limit of D}
\end{align}
To compare this result with the boundary-free result, we here give a self-consistent calculation. When there is no boundary, the vacuum polarization tensor $\Pi^{\mu\nu}$ takes the standard form,
\begin{equation}
    \Pi_{\mu\nu}(\scriptr)=\frac{2}{(4\pi)^2}\frac{N}{\scriptr^4}\mathcal{T}^{\mu\nu}(\scriptr),\quad \Pi_{\mu\nu}(k)=-\frac{Nk}{16}\left(\delta_{\mu\nu}-\frac{k_{\mu}k_{\nu}}{k^2}\right)\,.
\end{equation}
Under the same gauge condition $\partial^mD_{mn}=0$, we have the same Schwinger-Dyson equation about the transverse coordinates, although now transformed in the momentum space,
\begin{align}
   \int \frac{d^3k}{(2\pi)^3} D_{ml}(k)  \Pi_{ln}(k)e^{ik\cdot(r-r')}=-\left(\delta_{mn}-\frac{\partial_m\partial_n}{\partial_l\partial_l}\right)\delta(r-r').
\end{align}
Note that the $\sim k_lk_n$ part of $\Pi_{ln}$ contracted with $D_{ml}$ simply vanishes by the gauge condition, so we only keep the part $\sim \delta_{ln}$. Then tracing over indices gives
\begin{align}
    \int \frac{d^3k}{(2\pi)^3}(D_{\tau\tau}(k)+D_{xx}(k))\,  \frac{Nk}{16}\,e^{ik\cdot(r-r')}=\delta(r-r'),
\end{align}
and
\begin{equation}
    \mathfrak{D}(\scriptr)=D_{\tau\tau}(\scriptr)+D_{xx}(\scriptr) =\int \frac{d^3k}{(2\pi)^3} e^{-ik\cdot \scriptr}\frac{16}{Nk}=\frac{8}{N\pi^2}\frac{1}{\scriptr^2},
\end{equation}
which matches exactly Eq.~(\ref{eq: no boundary limit of D}).

\subsection{Boundary limit}
Although the full expression of $\mathfrak{D}=D_{\tau\tau}+D_{xx}$ is complicated in general, it simplifies a lot on the boundary, allowing one to solve each component of $D_{mn}$. 

By the conformal symmetry, one can assume the following general ansatz about $D_{mn}(\rho)$,
\begin{align}
    D_{mn}(\rho)=\rho_m\rho_n A(\rho)+\delta_{mn}B(\rho),
\end{align}
where $A(\rho)$ and $B(\rho)$ are unknown functions. Then by the gauge condition $\partial^mD_{mn}=0$ and the expression of $\mathfrak{D}(\rho)$,
\begin{align}
    &3A+\rho\frac{dA}{d\rho}+\frac{1}{\rho}\frac{dB}{d\rho}=0,\nonumber\\
   &\rho^2A+2B=\mathfrak{D}= \frac{4}{N\rho^2},
\end{align}
and the solution is 
\begin{align}
    A(\rho)=\frac{c+8\log \rho}{N\rho^4},\quad B(\rho)=\frac{4-c-8\log \rho}{2N\rho^2},
\end{align}
with an unknown constant $c$. We set $c=0$ because it breaks the rotational symmetry between $D_{\tau\tau}$ and $D_{xx}$. Thus,
\begin{align}
    \lim_{w_y,r_y'\to 0}D_{mn}(\rho)=\frac{1}{N}\left(\rho_m\rho_n \frac{8\log \rho}{\rho^4}+\delta_{mn}\frac{4-8\log \rho}{2\rho^2}\right).
\end{align}

\subsection{Full correlation function of the monopole current}
With the explicit expression of $\mathfrak{D}$, we derive the full correlation function of the gauge invariant quantity $f_{\mu\nu}$ (equivalently, the monopole current $J_{\mathrm{top}}^{\mu}$) in the bulk in this section, utilizing the residual conformal symmetry. 

We start with $J_{\mathrm{top}}^y=if_{\tau x}/2\pi$. The gauge fixing condition leads to the following relation,
\begin{align}
    \langle J_{\mathrm{top}}^y(r)J_{\mathrm{top}}^y(r')\rangle=-\frac{1}{(2\pi)^2}\langle f_{\tau x}(r)f_{\tau x}(r')\rangle=-\frac{1}{(2\pi)^2}(\partial_\tau\partial_{\tau'}+\partial_x\partial_{x'})\mathfrak{D},
\end{align}
so inserting the expression of $\mathfrak{D}$ from Eq.~(\ref{eq: D11+D22}) gives,
\begin{align}
    \langle J_{\mathrm{top}}^y(r)J_{\mathrm{top}}^y(r')\rangle=&-\frac{1}{N\pi^4\scriptr^4}\Bigg\{
    8-8v^4
    +(2v+2v^3)\Big[\pi^2+2\mathrm{Li}_2\left(\frac{(1-v)^2}{(1+v)^2}\right)-8\mathrm{Li}_2\left(\frac{1-v}{1+v}\right)\Big]
    \nonumber\\
    &+16v^2\log v+\frac{\rho^2}{\scriptr^2}\Bigg[
    -16 +8v^2-8v^4+16v^6-24(v^2+v^4)\log v\nonumber\\
    &    -(3v+3v^5+2v^3)\Big[\pi^2+2\mathrm{Li}_2\left(\frac{(1-v)^2}{(1+v)^2}\right)-8\mathrm{Li}_2\left(\frac{1-v}{1+v}\right)\Big]\Bigg]
 \Bigg\},
   \label{eq: Jy correlation function solved}
\end{align}
which has the boundary limit,
\begin{align}
    \lim_{y,y'\to0}\langle J_{\mathrm{top}}^y(\rho)J_{\mathrm{top}}^y(0)\rangle=\frac{4}{N\pi^2}\frac{1}{\rho^4},
\end{align}
recovering Eq.~(\ref{eq: Jtop correlator final}) of the main text.

The above result may be leveraged to solve a complete monopole current correlation function. The residual conformal symmetry in the bulk implies that the two-point correlation function of conserved currents takes the following form~\cite{McAvity:1995zd},
\begin{align}
     \langle J_{\mathrm{top}}^{\mu}(r)J_{\mathrm{top}}^{\nu}(r')\rangle
     =\frac{1}{\scriptr^4}\left(\mathcal{T}^{\mu\nu}(\scriptr)C(v)+X^{\mu}X'^{\nu}D(v)\right),
\end{align}
where 
\begin{align}
    X^{\mu}
    =v(\frac{2y}{\scriptr^2}\scriptr^{\mu}-\delta^{y\mu}),
    \qquad
    X'^{\mu}
    =v(-\frac{2y'}{\scriptr^2}\scriptr^{\mu}-\delta^{y\mu}),
\end{align}
and $C(v),D(v)$ are two scalar-valued functions of the crossing ratio $v$. Current conservation is encoded in the constraint,
\begin{align}
    v\frac{d}{dv}(C+D)=(d-1)D.
    \label{eq: CD ward identity}
\end{align}
Note that the deformation of the Ward identity to ${\partial^{\mu}J_{\mu}=\delta(y)J_y}$ means ${D(v\rightarrow1)}$ cannot vanish. 

Comparing $\langle J_{\mathrm{top}}^y(y)J_{\mathrm{top}}^y(r')\rangle$ in Eq.~(\ref{eq: Jy correlation function solved}) with this general form, we can identify 
\begin{align}
   N\pi^4(C+D)&=8-8v^4+16v^2\log v+(2v+2v^3)\Big[\pi^2+2\mathrm{Li}_2\left(\frac{(1-v)^2}{(1+v)^2}\right)-8\mathrm{Li}_2\left(\frac{1-v}{1+v}\right)\Big],\nonumber\\
  -N\pi^4(2C+(v^2+1)D) &=-16 +8v^2-8v^4+16v^6-24(v^2+v^4)\log v \nonumber\\
    &\quad\,-(3v+3v^5+2v^3)\Big[\pi^2+2\mathrm{Li}_2\left(\frac{(1-v)^2}{(1+v)^2}\right)-8\mathrm{Li}_2\left(\frac{1-v}{1+v}\right)\Big]\,.
\end{align}
The solution is 
\begin{align}
\label{eq: C final solution}
    C(v)=&-\frac{1}{N\pi^4}\frac{1}{v^2-1}\Bigg[8+\pi^2v-16v^2-2\pi^2v^3+16v^4+\pi^2v^5-8v^6+8v^2(1+v^2)\log v\nonumber\\
    &+v (v^2-1)^2\Big[2\mathrm{Li}_2\left(\frac{(1-v)^2}{(1+v)^2}\right)-8\mathrm{Li}_2\left(\frac{1-v}{1+v}\right)\Big]\Bigg],\\
    D(v)=&-\frac{1}{N\pi^4}\frac{v}{v^2-1}\Bigg[\pi^2+8v+2\pi^2v^2-24v^3-3\pi^2v^4+16v^5+8v(1-3v^2)\log v\nonumber\\
    &-(3v^2+1)(v^2-1)\Big[2\mathrm{Li}_2\left(\frac{(1-v)^2}{(1+v)^2}\right)-8\mathrm{Li}_2\left(\frac{1-v}{1+v}\right)\Big]\Bigg].
    \label{eq: D final solution}
\end{align}
One can check that the above $C(v)$ and $D(v)$ satisfy the Ward identity Eq.~(\ref{eq: CD ward identity}). Note that $C(v),D(v)\ge0$ in the physical range $v\in [0,1]$, which respects the reflection positivity in Euclidean space as expected~\cite{Herzog:2017xha}. 

The boundary condition $f_{\tau y}(y=0)=f_{xy}(y=0)=0$ is also respected, 
\begin{align}
     \lim_{y,y'\to0}\langle J_{\mathrm{top}}^m(\rho)J_{\mathrm{top}}^n(0)\rangle=\frac{1}{\rho^4}\mathcal{T}^{mn}(\rho)C(1)=0
\end{align}
here we used that $X_{m}=X'_{m}\equiv 0$ when $y=y'=0$. While $D(v=1)=4/N\pi^2$ is finite, which gives the expected residual of $\langle J_y(\rho)J_y(0)\rangle$ on the boundary. 

In the end, we remark that in the coincident limit where $\scriptr\to 0, v\to 0$, 
\begin{align}
    C(v=0)=\frac{8}{N\pi^4}, \quad  D(v=0)= 0,
\end{align}
and 
\begin{align}
    \lim_{\scriptr\to0} \langle J_{\mathrm{top}}^{\mu}(r)J_{\mathrm{top}}^{\nu}(r')\rangle=\frac{8}{N\pi^4}\frac{1}{\scriptr^4}\mathcal{T}^{\mu\nu}(\scriptr),
    \label{eq: final monopole current in position space}
\end{align}
which matches the monopole current correlation function in a boundary-free system in the large-$N$ limit~\cite{Huh:2013vga}.

\nocite{apsrev41Control}
\bibliographystyle{apsrev4-1}
\bibliography{boundaries}

\end{document}